\documentclass[aps,reprint,prd,superscriptaddress,nofootinbib,amsmath,amssymb]{revtex4-2}

\usepackage[utf8]{inputenc}
\usepackage{xcolor}
\usepackage{graphicx}
\usepackage{verbatim}
\usepackage{mathrsfs}
\usepackage{hyperref}

\newcommand{\rd}{\textrm{d}}
\newcommand{\ri}{\textrm{i}}
\newcommand{\NED}{\textrm{E}}
\newcommand{\sL}{\mathscr{L}}
\newcommand{\sH}{\mathscr{H}}
\newcommand{\sM}{\mathscr{M}}
\newcommand{\sP}{\mathscr{P}}
\newcommand{\sQ}{\mathscr{Q}}
\newcommand{\sF}{\mathscr{F}}
\newcommand{\sG}{\mathscr{G}}

\begin{document}

\title{Nonlinearly charging the conformally dressed black holes\texorpdfstring{\\}{} 
preserving duality and conformal invariance}

\author{Eloy \surname{Ay\'{o}n-Beato}}
\email[]{eloy.ayon-beato@cinvestav.mx}
\affiliation{Departamento de F\'{i}sica, CINVESTAV-IPN, A.P. 14-740, C.P. 07000, Ciudad de M\'exico, Mexico}

\author{Daniel \surname{Flores-Alfonso}}
\email[]{danflores@unap.cl}
\affiliation{Instituto de Ciencias Exactas y Naturales, Universidad Arturo Prat, Avenida Playa Brava 3256, 1111346, Iquique, Chile}
\affiliation{Facultad de Ciencias, Universidad Arturo Prat, Avenida Arturo Prat Chac\'on 2120, 1110939, Iquique, Chile}

\author{Mokhtar \surname{Hassaine}}
\email[]{hassaine@inst-mat.utalca.cl}
\affiliation{Instituto de Matem\'{a}ticas (INSTMAT), Universidad de Talca, Casilla 747, Talca 3460000, Chile}

\begin{abstract}
We start this paper by concisely rederiving ModMax, which is nothing but the unique nonlinear extension of Maxwell's equations preserving conformal and duality invariance. The merit of this new derivation is its transparency and simplicity since it is based on an approach where the elusive duality invariance is manifest. In the second part, we couple the ModMax electrodynamics to Einstein gravity with a cosmological constant together with a standard conformal scalar field, and new stationary spacetimes with dyonic charges are found. These solutions are later used as seed configurations to generate nonlinearly charged (super-)renormalizably dressed spacetimes by means of a known generating method that we extend to include any nonlinear conformal electrodynamics. We end by addressing the issue of how to generalize some of these results to include the recently studied non-Noetherian conformal scalar fields, whose equation of motion still enjoys conformal symmetry even though its action does not. It turns out that the static non-Noetherian conformally dressed black holes also become amenable to being charged by ModMax.
\end{abstract}

\maketitle

\section{Introduction}

It is certainly not an understatement to say that the paradigm of all effective low-energy field theories is due to the seminal works of Heisenberg, Euler, and Weisskopf \cite{Heisenberg:1936nmg,Weisskopf:406571}. In more modern language, it is agreed that by integrating out the degrees of freedom of the electron field in quantum electrodynamics (QED), one obtains an approximate description of the photon field deviating from the standard linear regime \cite{Dunne:2004nc}. This QED vacuum behavior makes nonlinear electrodynamics a semiphenomenological tool. As a result, several observational and experimental efforts have been deployed over the years to probe possible nonlinear effects in electrodynamics \cite{Ellis:2022uxv}. Well before these developments, Born and Infeld constructed a theory in which the electromagnetic field of a point charge is devoid of singularities \cite{Born:1934gh}. Indeed, by going beyond the principle of superposition, they were able to solve the problem of infinite self-energy, and they argued that there should be a maximum field strength similar to the upper limit provided by the speed of light in Special Relativity. This reasoning was found to be consistent with string theory decades later, where Born-Infeld theory reappears in the dynamics of strings and branes, and this maximum field strength is precisely associated with the fact that such extended objects cannot move faster than the speed of light \cite{Fradkin:1985qd,Bachas:1995kx}. This result further strengthens the interpretation of nonlinear electrodynamics as an effective field theory.

More recently, interest in nonlinear electrodynamics was considerably increased since it was shown that these theories not only tolerate the existence of finite self-energy solutions but actually allow more involved self-gravitating solutions free from any singularity \cite{Ayon-Beato:1998hmi,Ayon-Beato:1999qin,Ayon-Beato:1999kuh,Ayon-Beato:2004ywd}. In particular, the peculiar properties of these regular black hole solutions have motivated their study in the astrophysical settings by employing horizon-scale imaging of black holes \cite{Vagnozzi:2022moj}. Of course, almost all the involved electrodynamics become linear at the weak field limit, but all depart from the Maxwell behavior in a strong regime by exhibiting a characteristic length responsible for the regularization. 

The mentioned scale dependence is incompatible with conformal symmetry, one of the key features of standard Maxwell's theory in four dimensions and in fact, of almost the whole standard model. Hence, it is also desirable to have explicit self-gravitating configurations involving other electrodynamics that depart from the Maxwell one but still preserve the conformal invariance. It is worth mentioning that static black holes are known for electromagnetism theories with conformal symmetry for dimensions $D\not=4$, which can be achieved by considering as Lagrangian a power law of the single field strength invariant with a dimension-dependent exponent \cite{Hassaine:2007py}. These nonlinear models have also been shown to harmonize with conformally coupled scalar fields \cite{Cardenas:2014kaa}, even to the point of producing composite stealth configurations.\footnote{By stealth we mean nontrivial field configurations with a vanishing energy-momentum tensor; see Refs.~\cite{Ayon-Beato:2004nzi,Ayon-Beato:2005yoq,Ayon-Beato:2015qfa,Ayon-Beato:2015mxf,Ayon-Beato:2024xgp}} It is natural to consider scalar fields at the same time since they are precisely the simplest way to realize conformal symmetry. In this respect, we mention that the most general second-order conformally invariant scalar field equation in four dimensions arising from an action principle was recently found in \cite{Ayon-Beato:2023bzp} after the precursor results of Ref.~\cite{Fernandes:2021dsb}. 

In the context of black holes the relevance of conformally invariant fields has been evident since the pioneering and independent works of Refs.~\cite{Bocharova:1970skc,Bekenstein:1974sf}, where the first example of a black hole dressed by a conformal scalar field was provided. This solution turns out to be pathological due to the divergence of the scalar field at the event horizon \cite{Sudarsky:1997te}. Remarkably, this pathology can be cured just by turning on a cosmological constant together with the unique conformally invariant potential in four dimensions \cite{Martinez:2002ru}. This spherical black hole allows an electrically charged extension \cite{Martinez:2002ru} and an analog whose event horizon is negatively curved \cite{Martinez:2005di}. For completeness we also mention that they have been generalized in several ways as solutions; see, e.g., \cite{Anabalon:2012tu,Bardoux:2013swa,Ayon-Beato:2015ada,Barrientos:2022avi}. Of course, the dyonic extension of this so-called charged MTZ black hole is straightforward thanks to the duality and conformal symmetries of the standard Maxwell theory. Outstandingly, it turns out that Maxwell electrodynamics is not the only one that shares such properties. In fact, there exists a unique nonlinear extension of the Maxwell theory that inherited both its conformal and duality symmetry, which was recently proposed in Ref.~\cite{Bandos:2020jsw} and dubbed \emph{ModMax}. The main goal of this paper is to explore if the mere assumption of these two ingredients, i.e.\ duality as well as conformal invariance, is enough to find further generalizations of the known conformally dressed black holes. The results of our exploration are positive and, in what follows we summarize the different steps to achieve our objective. 

The plan of the paper is organized as follows. In order to be self-contained but also for usefulness, we start in Sec.~\ref{sec:NLE} by providing a survey of nonlinear electrodynamics {\it \`a la } Salazar, Garc\'{\i}a and Pleba\'nski \cite{Salazar:1987ap}. This step is needed since it allows a transparent definition of duality-invariant theories that we will carefully justify in Sec.~\ref{subsec:DS}. The conformal symmetry is then characterized in Sec.~\ref{subsec:CI} and later used beside the previous results to allow a clear rederivation of the ModMax theory in Sec.~\ref{subsec:MMd}. In Sec.~\ref{sec:MMchargedMTZ}, we provide the dyonic generalization, supplemented with a nontrivial NUT parameter, of the MTZ black hole nonlinearly charged through the ModMax theory. Since we go beyond Maxwell theory respecting power-counting renormalizability it makes total sense to do something similar with the scalar field. Fortunately, in Ref.~\cite{Ayon-Beato:2015ada} it was shown by means of a generating technique that the charged MTZ solution allows a generalization supported by a (super-)renormalizable self-interaction potential. Here, we first extend this generating technique to include any conformal nonlinear electrodynamics and then construct new (super-)renormalizably dressed spacetimes in Sec.~\ref{sec:SR}. This work is finally completed by adding to the scalar field a term that breaks its conformal invariance at the level of the action but not at the level of its equation of motion, i.e.\ we also address the study of the so-called non-Noetherian conformal scalar fields \cite{Fernandes:2021dsb,Ayon-Beato:2023bzp}. In Sec.~\ref{sec:nonNoether} we succeed to charge with ModMax the static non-Noetherian conformally dressed black holes originally found by Fernandes in Ref.~\cite{Fernandes:2021dsb}. The last section, \ref{sec:Conclu}, is devoted to our conclusions.

%%%%%%%%%%%%%%%%%%%%%%%%%%%%%%%%%%%%%%%%%%%%%%%%%%
\section{Nonlinear electrodynamics\label{sec:NLE}}
%%%%%%%%%%%%%%%%%%%%%%%%%%%%%%%%%%%%%%%%%%%%%%%%%%

As is defined in the Pleba\'nski book \cite{Plebanski:1970}, nonlinear electrodynamics are described by the following action principle {\small
\begin{equation}
S_{\NED}[g,A,P]=-\frac{1}{4\pi}\!\int\!\rd^4x \sqrt{-g}
\!\left(\!\frac{1}{2}F_{\mu\nu}P^{\mu\nu}-{\sH}(\sP,\sQ)\!\right)\!, \label{SNED}
\end{equation}}%
which depends on the metric $g_{\mu\nu}$, the gauge potential $A_\mu$ and the antisymmetric tensor $P^{\mu\nu}$. Here the structural function $\sH$ describes the precise nonlinear electrodynamics and depends, in general, on the two Lorentz scalars that can be constructed with $P^{\mu\nu}$ \cite{Born:1934gh}; see the first equality of Eq.~\eqref{parabolicP}. As usual, the field strength is related to the gauge potential as $F=\rd A$, ensuring the Faraday equations
\begin{equation}
 \rd F=0. \label{Bianchi}
\end{equation}
On the other hand, the variation of action \eqref{SNED} with respect to the gauge potential leads to the Maxwell equations
\begin{equation}
 \rd\star P=0, \label{Maxwell}
\end{equation}
where $\star$ stands for the Hodge dual, whereas varying \eqref{SNED} with respect to the antisymmetric tensor $P^{\mu\nu}$ yields the constitutive relations
\begin{equation}
 F_{\mu\nu}={\sH}_{\sP}P_{\mu\nu} + {\sH}_{\sQ}\star P_{\mu\nu}, \label{CR}
\end{equation}
with the subindices of $\sH$ indicating partial differentiation, e.g., ${\sH}_{\sP}={\partial\sH}/{\partial\sP}$. Notice that Maxwell electrodynamics is recovered for $\sH=\sP$, giving linear constitutive relations. Lastly, the corresponding energy-momentum tensor reads
\begin{equation}
4\pi T^{\NED}_{\mu\nu} = F_{\mu\alpha}{P_{\nu}}^{\alpha}
- g_{\mu\nu}\left( \frac12F_{\alpha\beta}P^{\alpha\beta} - {\sH} \right). \label{TEmunu}
\end{equation}
The main motivation for the action principle \eqref{SNED} is that now Maxwell equations \eqref{Maxwell} remain linear as the Faraday ones \eqref{Bianchi} while the nonlinearity is encoded into the constitutive relations \eqref{CR}. Consequently, Maxwell equations \eqref{Maxwell} can be now understood just like the Faraday ones \eqref{Bianchi}, i.e., implying the local existence of a vector potential $\star P=\rd A^*$. Therefore, from the point of view of the action principle \eqref{SNED}, a solution to nonlinear electrodynamics can be understood as a pair of vector potentials $A$ and $A^*$ compatible with the constitutive relations \eqref{CR}. Additionally, since in four dimensions both Faraday \eqref{Bianchi} and Maxwell \eqref{Maxwell} equations define conservation laws, there are conserved quantities related to them defined by the following integrals 
\begin{equation}
p=\frac{1}{4\pi}\int_{\partial\Sigma} F, \qquad 
q=\frac{1}{4\pi}\int_{\partial\Sigma} \star P, \label{charges}
\end{equation}
where the integration is taken at the boundary of constant time hypersurfaces $\Sigma$; obviously, these are nothing other than the magnetic and electric charges, respectively.

After this brief and useful introduction, and in order to prepare for what follows, we review a strategy that has proved to be fruitful when nonlinear electrodynamics is considered in General Relativity \cite{Plebanski:1970}. This strategy simply consists of working in a null tetrad of the spacetime metric 
\begin{equation}
g=2e^1\otimes_\text{s} e^2+2e^3\otimes_\text{s} e^4,
\end{equation}
aligned along the common eigenvectors of the electromagnetic fields, i.e.\
\begin{equation}
F+\ri\star P=(D+\ri B) e^1\wedge e^2+(E+\ri H) e^3\wedge e^4. \label{aligned}
\end{equation}
Here, the first pair of the tetrad is composed of complex conjugates one-forms, while the last pair is real. Additionally, it has been implicitly assumed that the electromagnetic configuration is algebraically general; namely, the real invariants $E$, $B$, $D$, and $H$ which are related to the eigenvalues are not all zero at the same time. We also remark that the scalars $E$ and $D$ are associated with the intensity of the electric field and electric induction, respectively, as perceived in the null frame, while $H$ and $B$ are their magnetic counterparts. In terms of the aligned tetrad invariants \eqref{aligned}, the standard invariants take the form {\small
\begin{subequations} \label{parabolic}
\begin{align}
\sF+\ri\sG&\equiv\frac{1}{4}F_{\mu\nu}F^{\mu\nu}+\frac{\ri}{4}F_{\mu\nu}\star F^{\mu\nu}=-\frac{1}{2}(E+\ri B)^2 \label{parabolicF}, \\
\sP+\ri\sQ&\equiv\frac{1}{4}P_{\mu\nu}P^{\mu\nu}+\frac{\ri}{4}P_{\mu\nu}\star P^{\mu\nu}=-\frac{1}{2}(D+\ri H)^2, \label{parabolicP}
\end{align}
\end{subequations}}%
resulting in a parabolic relation between them. Therefore, the structural function is reparameterized as $\sH(\sP,\sQ)=\sH(D,H)$, which leads to a simpler version of the constitutive relations \eqref{CR} that now reads
\begin{equation}
E+\ri B=(-\partial_D+\ri\partial_H)\sH. \label{CRH}
\end{equation}
This is not the only advantage of choosing an aligned tetrad, it also results in a diagonalization of the energy-momentum tensor allowing only two independent components. The latter are better expressed through the trace, $\text{tr}\,T^{\NED}$, of \eqref{TEmunu} together with its traceless part, $\hat{T}^{\NED}\equiv T^{\NED}-\tfrac14g\,\text{tr}\,T^{\NED}$, according to
\begin{subequations}\label{energymomentumH}
\begin{align}
2\pi\,\text{tr}\,T^{\NED}&=DE-BH+2\sH, \label{traceH}\\
4\pi\hat{T}^{\NED}&=(DE+BH)(e^1\otimes_\text{s}e^2-e^3\otimes_\text{s}e^4). \label{tracelessH}
\end{align}
\end{subequations}
Here $E$ and $B$ must be determined from the constitutive relations \eqref{CRH}. This approach has been employed since the seminal work of Pleba\'nski \cite{Plebanski:1970} to the formulations of nonlinear electrodynamics pioneered in Ref.~\cite{Salazar:1987ap} which we also review below. The power of this approach is such that it has been instrumental in the derivation of the first genuine example of a spinning nonlinearly charged black hole \cite{Garcia-Diaz:2021bao,Garcia-Diaz:2022jpc,Ayon-Beato:2022dwg}.

The action principle \eqref{SNED} is concretely obtained as a Legendre transform from a Lagrangian, $-\tfrac1{4\pi}\sL(\sF,\sG)$, which prompts dubbing the action structural function $\sH(\sP,\sQ)$ as the ``Hamiltonian''. This total Legen\-dre transform is concretely given in terms of the new variables by
\begin{equation}
\sL(E,B)=BH-DE-\sH(D,H).\label{L}
\end{equation}
This highlights that in the more known Lagrangian formulation the fundamental variables are instead $E$ and $B$, where the others must be determined from the constitutive relations. Using \eqref{CRH} as $\rd\sH=-E\rd D+B\rd H$ in the differential of \eqref{L}, the constitutive relations acquire now the alternative simple form
\begin{equation}
D+\ri H=(-\partial_E+\ri\partial_B)\sL. \label{CRL}
\end{equation}
Furthermore, the total Legendre transform \eqref{L} motivates the definition of the following two partial Legendre transforms
\begin{subequations}\label{M+-}
 \begin{align}
  \sM^+(D,B)&=BH-\sH(D,H), \label{M+}\\
  \sM^-(E,H)&=DE+\sH(D,H), \label{M-}
 \end{align}
\end{subequations}
which were first introduced in Ref.~\cite{Salazar:1987ap} and led to two alternative dual descriptions of nonlinear electrodynamics, using purely inductions or intensities as independent variables. As was first pointed out and exploited in \cite{Salazar:1987ap}, these dual formulations are precisely the ideal ones to transparently describe theories invariant under duality rotations, a fact that will be reviewed in Sec.~\ref{subsec:DS}. Additionally, they are indispensable to determine the electrodynamics supporting the spinning nonlinearly charged black holes of \cite{Garcia-Diaz:2021bao,Garcia-Diaz:2022jpc} as was shown in \cite{Ayon-Beato:2022dwg}. Correspondingly, the constitutive relations in the mixed representations are written as
\begin{subequations}\label{CRM}
\begin{align}
  E+\ri H=(\partial_D+\ri\partial_B)\sM^+, \label{CRM+}\\
  D+\ri B=(\partial_E+\ri\partial_H)\sM^-. \label{CRM-}
\end{align}
\end{subequations}
Regarding the energy-momentum tensor, its traceless part is formally written again as in \eqref{tracelessH} but considering the new constitutive relations above, while the trace changes according to the partial Legendre transforms \eqref{M+-}
\begin{equation}
2\pi\,\text{tr}\,T^{\NED}=\pm(DE+BH-2\sM^{\pm}). \label{trace}
\end{equation}
After clearly defining duality-invariant theories through these formalisms in the next section, we shall apply the above expression to determine their zero trace subclass defining the ModMax theory.

%%%%%%%%%%%%%%%%%%%%%%%%%%%%%%%%%%%%%%%%%%%%%%%%%%%%%%%%%%
\section{A simpler derivation of the ModMax electrodynamics}
%%%%%%%%%%%%%%%%%%%%%%%%%%%%%%%%%%%%%%%%%%%%%%%%%%%%%%%%%%

In this section, we follow the formalism of the previous section to rederive the ModMax theory \cite{Bandos:2020jsw} in a novel and comprehensive way. Our approach will be based on the inductions and intensities formulations with structural functions $\sM^+$ and $\sM^-$, respectively, which in turn provide a direct and transparent way of deriving the theory from its defining symmetries. A symmetry-based approach was previously undertaken in \cite{Kosyakov:2020wxv} but employed a Lagrangian description. In contrast with the $\sM^\pm$ formulations, the resulting condition is nonlinear as we will explain later, and it allows no explicit general solution as recently emphasized in \cite{Russo:2024llm}.

%%%%%%%%%%%%%%%%%%%%%%%%%%%%%%%%%%%%%%%%%%%%%%%
\subsection{Duality symmetry\label{subsec:DS}}
 %%%%%%%%%%%%%%%%%%%%%%%%%%%%%%%%%%%%%%%%%%%%%%%

The main challenge is how to implement the duality symmetry. Fortunately, this problem was cleverly solved by Salazar, Garc\'{\i}a and Pleba\'nski in \cite{Salazar:1987ap}. We didactically follow their approach to make evident its advantages. Duality symmetry is related to the possibility of interchanging the Faraday \eqref{Bianchi} and Maxwell \eqref{Maxwell} equations, which are preserved by the more general rotations
\begin{equation}
\tilde{F}+\ri\star\tilde{P}=e^{\ri\varphi}(F+\ri\star P). \label{duality-rotationF*P}
\end{equation}
From the conserved charges \eqref{charges}, these rotations are a covariant nonlinear realization of the well-known electric-magnetic duality
\begin{equation}
\tilde{p}+\ri\tilde{q}=e^{\ri\varphi}(p+\ri q). \label{electric-magnetic-duality}
\end{equation}
However, this is not the end of the story since the involved fields are not independent as they are tied by the constitutive relations \eqref{CR}. The circumstances under which the latter also remain unchanged after the rotations above are far from obvious, but we will show they are simpler to analyze in the $\sM^\pm$ formalisms. In fact, as we plan to couple the resulting duality-invariant nonlinear electrodynamics to General Relativity, we first demand the invariance of the energy-momentum tensor and show later that the emerging duality conditions are enough to also ensure preserving the constitutive relations.

Using an aligned tetrad \eqref{aligned}, the duality rotations \eqref{duality-rotationF*P} becomes
\begin{subequations}\label{duality-rotation}
\begin{align}
\tilde{D}+\ri\tilde{B}&=e^{\ri\varphi}(D+\ri B), \\
\tilde{E}+\ri\tilde{H}&=e^{\ri\varphi}(E+\ri H),
\end{align}
\end{subequations}
while their infinitesimal version 
\begin{subequations}
\begin{align}
\tilde{D}+\ri\tilde{B}&=D+\ri B+\varphi\,X(D+\ri B)+\cdots, \\
\tilde{E}+\ri\tilde{H}&=E+\ri H+\varphi\,X(E+\ri H)+\cdots,
\end{align}
\end{subequations}
defines the generator of duality rotations in this representation 
\begin{equation}
X=-B\partial_D+D\partial_B-H\partial_E+E\partial_H. \label{generator}
\end{equation}
As previously announced, we shall demand now the invariance of the energy-momentum tensor under duality rotations, which infinitesimally requires the generator \eqref{generator} to preserve their two independent components. The traceless part \eqref{tracelessH} is already duality invariant since its single component identically satisfies
\begin{equation}
X(DE+BH)=0.
\end{equation}
Then the duality invariance of the trace \eqref{trace} reduces to the following conditions
\begin{equation}
X(\sM^\pm)=0. \label{X(M)=0}
\end{equation}
These last requirements are enough to also preserve the constitutive relations \eqref{CRM} under duality rotations since the corresponding infinitesimal conditions can be rewritten as {\small 
\begin{align*}
X\left(E+\ri H-[\partial_D+\ri\partial_B]\sM^+\right)=-
(\partial_D+\ri\partial_B)X(\sM^+)&=0,\\
X\left(D+\ri B-[\partial_E+\ri\partial_H]\sM^-\right)=-(\partial_E+\ri\partial_H)X(\sM^-)&=0.
\end{align*}}%
Notice that if one would have started by analyzing first the invariance of the constitutive relations, one would have obtained from the previous expressions the preliminary conditions $X(\sM^\pm)=\text{const.}$, which is exactly equivalent by \eqref{CRM} to the preliminary duality condition $DH-BE=\text{const.}$ obtained in Ref.~\cite{Gibbons:1995cv}. Anyway, the later restriction on the energy-momentum tensor, also enforced in \cite{Gibbons:1995cv}, finally imposes the vanishing of this constant. The advantage here is that the duality condition \eqref{X(M)=0} is expressed as a constraint on the structural functions which can be phrased as ``a nonlinear electrodynamics is duality invariant if and only if its structural functions $\sM^+(D,B)$ and $\sM^-(E,H)$ are preserved by duality rotations.'' This is equivalent to the following homogeneous linear first-order PDE {\small 
\begin{equation}
(-B\partial_D+D\partial_B)\sM^+=0, \quad
(-H\partial_E+E\partial_H)\sM^-=0,
\label{dual}
\end{equation}}%
whose general solutions are, respectively, of the form {\small
\begin{equation}
\sM^+=\sM^+\left(\frac{D^2+B^2}2\right), \quad
\sM^-=\sM^-\left(\frac{E^2+H^2}2\right). \label{dualM+-}
\end{equation}}%
Hence, as was first realized in \cite{Salazar:1987ap}, a generic duality-invariant nonlinear electrodynamics is necessarily described by the above structural functions, each generally depending on a single rotation-invariant variable built with the corresponding aligned invariants.

We end this subsection by a last comment regarding the duality condition $BE=DH$ in the Lagrangian and Hamiltonian formulations. Using the constitutive relations \eqref{CRL} or \eqref{CRH}, this duality requisite can be expressed in either of the following two ways
\begin{equation}
\partial_E\sL\partial_B\sL=-EB,\qquad 
\partial_D\sH\partial_H\sH=-DH. \label{nl1oPDE}
\end{equation}
These are nonlinear first-order PDE of Hamilton-Jacobi type \cite{Gibbons:1995cv}, and it is a nontrivial task to find their general solutions. In the case of the above equations, it is possible to find their general solutions depending on an arbitrary function; see Ref.~\cite{Ayon-Beato:2024xgp} for a recent account of the several methods used to achieve this task. Unfortunately, as occurs for all nonlinear first-order PDE, such general solutions are only implicitly given (see the Appendix and the recent Ref.~\cite{Russo:2024llm}). This makes impossible a transparent treatment of duality-invariant theories in the conventional formulations, i.e., the occurrence or not of duality invariance needs to be checked case by case.

%%%%%%%%%%%%%%%%%%%%%%%%%%%%%%%%%%%%%%%%%%%%%%%%%%%%
\subsection{Conformal invariance\label{subsec:CI}}
%%%%%%%%%%%%%%%%%%%%%%%%%%%%%%%%%%%%%%%%%%%%%%%%%%%%

We now turn to conformal invariance, the other defining symmetry of the ModMax electrodynamics. It is easier to implement since it is well-known that conformal symmetry necessarily implies a traceless energy-momentum tensor. Hence, we must set the trace \eqref{trace} to zero, which by taking into account the constitutive relations \eqref{CRM} yields the following inhomogeneous linear first-order PDE
\begin{subequations}\label{conformal}
\begin{align}
(D\partial_D+B\partial_B)\sM^+&=2\sM^+, \\
(E\partial_E+H\partial_H)\sM^-&=2\sM^-.
\end{align}
\end{subequations}
These linear PDE are easily solved and the result (which can also be obtained in the other formulations) is that the structural functions characterizing all the conformally invariant nonlinear electrodynamics are necessarily homogeneous. For example, the degree of the previous functions $\sM^\pm$ must be two, and by directly taking the trace of \eqref{TEmunu} it follows that the degree of $\sH(\sP,\sQ)$ must be one. In order to exhibit that such conditions are enough to guarantee the conformal invariance of action \eqref{SNED}, we explicitly write down the conformal weights of its different ingredients 
\begin{equation}
(g_{\mu\nu},A_\mu,P^{\mu\nu})\,\mapsto\,
(\Omega^2g_{\mu\nu},A_\mu,\Omega^{-4}P^{\mu\nu}).
\end{equation}
They, together with the degree one homogeneity of the Hamiltonian, imply the latter transforms with the appropriate weight, i.e., {\small 
\begin{equation}
\sH(\sP,\sQ)\mapsto
\sH(\Omega^{-4}\sP,\Omega^{-4}\sQ)=
\Omega^{-4}\sH(\sP,\sQ),
\end{equation}}%
ensuring in this way the required conformal invariance of action \eqref{SNED}.

%%%%%%%%%%%%%%%%%%%%%%%%%%%%%%%%%%%%%%%%%%%%%%%%
\subsection{ModMax derivation\label{subsec:MMd}}
%%%%%%%%%%%%%%%%%%%%%%%%%%%%%%%%%%%%%%%%%%%%%%%%

After having transparently characterized both duality and conformal symmetries, it will turn out to easily derive the ModMax theory \cite{Bandos:2020jsw}. The single-argument dependence of duality-invariant theories \eqref{dualM+-} reduces the homogeneity defining PDE \eqref{conformal}, warranting conformal symmetry to the simple ordinary equations
\begin{subequations}\label{MMordinary}
\begin{align}
\frac12\left(D^2+B^2\right){\sM^+}'&=\sM^+,\\
\frac12\left(E^2+H^2\right){\sM^-}'&=\sM^-,
\end{align}
\end{subequations}
whose general solutions are given by
\begin{subequations}\label{ModMaxM+-}
\begin{align}
\sM^+_\text{MM}(D,B)&=\frac{e^{-\gamma}}2\left(D^2+B^2\right),\\
\sM^-_\text{MM}(E,H)&=\frac{e^{\gamma}}2\left(E^2+H^2\right).
\end{align}
\end{subequations}
These structural functions define the so-called ModMax electrodynamics, i.e., the generic family of duality-invariant theories that enjoy conformal symmetry at the same time. The difference between the above integration constants is due to the fact that it is enough to integrate only one of the structural functions, since the other is obtained from the Legendre transforms \eqref{M+-} and results in the inverse integration constant of the former. In order to write the theory in a more standard form, notice that due to the squared dependence \eqref{ModMaxM+-} the resulting constitutive relations \eqref{CRM} become linear and trivially invertible
\begin{equation}
E+\ri H=e^{-\gamma}(D+\ri B).
\end{equation}
That turns into a straightforward task to obtain the other structural functions from the Legendre transforms \eqref{M+-} and \eqref{L}, also yielding a squared dependence in terms of the aligned invariants
\begin{subequations}\label{ModMaxLHp}
\begin{align}
\sL_\text{MM}(E,B)&=-\frac12\left(e^{\gamma}E^2
-e^{-\gamma}B^2\right),\label{L_MM(E,B)}\\
\sH_\text{MM}(D,H)&=-\frac12\left(e^{-\gamma}D^2
-e^{\gamma}H^2\right).\label{H_MM(D,H)}
\end{align}
\end{subequations}
We are now ready to express the Lagrangian and Hamiltonian in terms of the conventional invariants by inverting the parabolic relations \eqref{parabolic} between the two families of invariants. This finally allows us to write down the standard structural functions for ModMax theory in their best known form
\cite{Bandos:2020jsw}
\begin{subequations}\label{ModMaxLH}
\begin{align}
\sL_\text{MM}(\sF,\sG)&=\cosh(\gamma)\sF
-\sinh(\gamma)\sqrt{\sF^2+\sG^2},\\
\sH_\text{MM}(\sP,\sQ)&=\cosh(\gamma)\sP
+\sinh(\gamma)\sqrt{\sP^2+\sQ^2}.\label{BLST}
\end{align}
\end{subequations}
This is the precise family of electrodynamics we shall use in the rest of the paper.

%%%%%%%%%%%%%%%%%%%%%%%%%%%%%%%%%%%%%%%%%%%%%%%%%%%%%%%%
\section{Nonlinearly charging the MTZ black hole via ModMax
\label{sec:MMchargedMTZ}}
%%%%%%%%%%%%%%%%%%%%%%%%%%%%%%%%%%%%%%%%%%%%%%%%%%%%%%%%

We now proceed to charge the conformally dressed MTZ black holes \cite{Martinez:2002ru,Martinez:2005di} beyond the linear regime but still preserving the conformal and duality symmetries of its involved sources. In concrete terms, this means considering the ModMax nonlinear electrodynamics \cite{Bandos:2020jsw} instead of Maxwell, and hence dealing with the following action principle 
\begin{subequations} {\small 
\begin{equation}
S[g,\Phi,A,P] = \int \rd^{4}x \sqrt{-g} \!\left(\! \frac{R-2\Lambda}{2\kappa} + L_\text{CS}+L_\text{MM}\!\right)\!, \label{action}
\end{equation}}%
where the involved matter Lagrangian densities
\begin{align}
L_\text{CS}&=- \frac{1}{2}\nabla_{\mu}\Phi\nabla^{\mu}\Phi -\frac{1}{12} R \Phi^2 - \lambda \Phi^{4},\label{L_CS} \\
-4\pi L_\text{MM}&=\frac{1}{2}F_{\mu\nu}P^{\mu\nu}-\sH_\text{MM}(\sP,\sQ),
 \label{LNED}
\end{align}
\end{subequations}
describe a conformal scalar field and the ModMax nonlinear electrodynamics, respectively. The structural function \eqref{BLST} of the last theory fixes the constitutive relations \eqref{CR} to
\begin{subequations}\label{eom} {\small 
\begin{equation}
F=\left(\cosh(\gamma)+\frac{\sinh(\gamma)\sP}  {\sqrt{\sP^2+\sQ^2}}\right)P
+\frac{\sinh(\gamma)\sQ}{\sqrt{\sP^2+\sQ^2}}\star P, \label{modmax}
\end{equation}}%
and since the role of Faraday \eqref{Bianchi} and Maxwell \eqref{Maxwell} equations is to define a pair of vector potentials $F=\rd A$ and $\star P=\rd A^*$, the previous relations are the only restriction these vector fields must satisfy. The remaining relevant variations are those with respect to the conformal scalar field and the metric tensor yielding  
\begin{align}
\Box \Phi -\frac{1}{6} R\Phi &= 4\lambda\Phi^{3},
\label{Eq_CS} \\
G_{\mu\nu} + \Lambda g_{\mu\nu} &= \kappa \left(T^\text{CS}_{\mu\nu}+T^\text{MM}_{\mu\nu}\right),
 \label{EFE}
\end{align}
respectively, where both conformal contributions to the energy-momentum tensor are given by {\small 
\begin{align}
T^\text{CS}_{\mu\nu} ={}& \nabla_{\mu}\Phi\nabla_{\nu}\Phi - g_{\mu\nu} \left( \frac{1}{2}\nabla_{\alpha}\Phi\nabla^{\alpha}\Phi + \lambda\Phi^{4} \right) \notag\\
 &+ \frac{1}{6}\left( g_{\mu\nu}\Box - \nabla_{\mu}\nabla_{\nu} + G_{\mu\nu} \right)\Phi^2,\label{Tmunu}\\
4\pi T^\text{MM}_{\mu\nu} \!={}&\! \!\left(\!\cosh(\gamma)+\frac{\sinh(\gamma)\sP}  {\sqrt{\sP^2+\sQ^2}}\!\right)\!
(P_{\mu\alpha}{P_{\nu}}^{\alpha} - \sP g_{\mu\nu}), \label{TMMmunu}
\end{align}}%
\end{subequations}
and the last expression is obtained from the generic energy-momentum tensor for a nonlinear electrodynamics \eqref{TEmunu}, after substituting the ModMax structural function \eqref{BLST}.

We are interested in constructing at once stationary solutions extended with a Taub-NUT parameter. This goal is motivated by the fact that, on the one hand, the NUT solution generalizing the conventionally charged conformally dressed black hole is already known \cite{Bardoux:2013swa}, and on the other, we can exploit the fact that the NUT solutions were generically obtained for all duality-invariant nonlinear electrodynamics in the absence of the scalar field \cite{Salazar:1987ap}. Regarding the latter, the precise case of ModMax was recently reanalyzed in Refs.~\cite{BallonBordo:2020jtw,Flores-Alfonso:2020nnd}. It is easy to incorporate a constant conformal scalar field on top of the previous configurations \cite{Zhang:2021qga}, but the real challenge is to achieve the goal with a nontrivial scalar field.  All the previously mentioned configurations are obviously inspired by the vacuum spacetime solutions originally found and characterized in Refs.~\cite{Taub:1950ez,Newman:1963yy,Misner:1963fr}. Starting from this observation, we will consider the following generic Taub-NUT ansatz for our task
\begin{subequations}\label{Taub-NUTansatz}
\begin{align}
\rd s^2 ={}& -f(r)\left(\rd t +\frac{4n}k\sin^2{(\sqrt{k}\theta/2)}\rd\phi\right)^2
 \notag\\
 & + \frac{\rd r^2}{f(r)} + (r^2+n^2)\rd\Omega_k^2, \label{metricAnsatz}\\
\Phi ={}&\Phi(r),\frac{}{}\\
A ={}& A_t(r) \left(\rd t +\frac{4n}k\sin^2{(\sqrt{k}\theta/2)}\rd\phi\right),\\
A^{*} ={}& A^{*}_t(r) \left(\rd t +\frac{4n}k\sin^2{(\sqrt{k}\theta/2)}\rd\phi\right),
\end{align} 
where the two-dimensional geometry 
\begin{equation}
\rd\Omega_k^2=\rd\theta^2 + \frac1k{\sin^2{(\sqrt{k}\theta)}}\rd\phi^2,
\end{equation}
\end{subequations}
describes a base space of constant curvature $k$, and $n$ denotes the NUT parameter. 

The consistent self-gravitating solution of the conformal composite described by Eqs.~\eqref{eom} for the above Taub-NUT ansatz is given by
\begin{subequations} \label{solution}
\begin{align}
  f(r) &= -\frac{\Lambda}{3}(r^2+n^2)+
\frac{\left(k-\frac43\Lambda n^2\right)\left(r-m\right)^2}{r^2+n^2}, \label{metricFunction}
 \\
\Phi(r) &= \sqrt{-\frac{\Lambda}{6\lambda}(m^2+n^2)}
\frac1{r-m},\label{Phi} \\
 A_t(r) &= h(r;p,-q), \\
 A^{*}_t(r) &= h(r;q,p),
\end{align}
where the same family of functions
\begin{align}
h(r;q,p)\equiv{}&\frac{q}{2n}\cos\left[ 
2e^{-\gamma}\arctan\left( \frac{n}{r} \right) \right]
\notag\\
&+\frac{p}{2n}\sin\left[ 
2e^{-\gamma}\arctan\left( \frac{n}{r} \right) \right],
\label{h}
\end{align}
determines the vector potentials with different parameterizations. Here, $m$, $q$, and $p$ are integration constants that, together with $n$, are tied via the coupling constants of the theory through the following relation
\begin{align}
e^{-\gamma}(q^2+p^2)={}&\frac{2\pi}{9}
\!\left( k-\frac{4}{3}\Lambda n^2 \right)\notag \\ &\times\left(\frac{36}{\kappa}+\frac{\Lambda}{\lambda}\right)\!(m^2+n^2).
\label{rest}
\end{align}
\end{subequations}

A careful examination of the solution reveals that, in units where $\kappa=8\pi$, $m$ is exactly the mass of the spherical black holes resulting at $k=1$ in the static limit $n=0$ we discuss later. This can be calculated using the successful off-shell formalism proposed in \cite{Kim:2013zha} and discussed in detail in \cite{Ayon-Beato:2015jga}; see also \cite{Ayon-Beato:2019kmz}. Besides, $q$ and $p$ are nothing but the electric and magnetic charges, respectively, which is explicitly checked also in the spherical case $k=1$ by integrating the right-hand sides of \eqref{charges} as usual over a sphere at infinity. 

The electric-magnetic duality is explicitly manifested in the solution since the change 
\begin{equation}
(p,q)\mapsto(-q,p),
\end{equation}
maps $A^{*}$ into $A$ preserving the parametric constraint \eqref{rest}, which interchanges the roles of the magnetic and electric charges \eqref{charges}. On the other hand, the metric and the scalar field are unaffected by the duality transformation since they are independent of the electromagnetic charges. Something similar applies to the full duality rotations \eqref{electric-magnetic-duality}, which now rotates the vector potentials and the result is that the solution \eqref{solution} is mapped to itself but with magnetic charge $\tilde{p}$ and electric charge $\tilde{q}$.

As a comment regarding (anti-)self-duality, it can be achieved as usual in the Euclidean continuation, which is obtained by means of the following Wick rotations 
\begin{equation}
(t,n,q,A^*)\rightarrow(\ri t,\ri n,\ri q,\ri A^*),
\end{equation}
giving for the Euclidean vector potentials $A_t(r)=h(r;p,q)$ and $A^*_t(r)=h(r;q,p)$. The (anti-)self-dual points are defined by $m=\pm n$ and $p=\pm q$, yielding $A=\pm A^*$ with vanishing electromagnetic energy-momentum tensor and making the scalar field trivial, thus reducing the Euclidean metric to a vacuum configuration with (anti-)self-dual curvature, in agreement with previous results \cite{BallonBordo:2020jtw,Flores-Alfonso:2020nnd,Barrientos:2022yoz,Colipi-Marchant:2023awk}.

Before continuing with the characterization of the solutions, we want to ponder the meaning of the constraint \eqref{rest}. Indeed, it has long been known that the Weyl tensor decomposition in terms of electric and magnetic parts in the paradigm of Taub-NUT spacetimes provides gravitoelectric and gravitomagnetic interpretations for the mass $m$ and the NUT parameter $n$, respectively \cite{Demianski:1966}. In this sense the dyonic character of solutions like \eqref{solution} is twofold, since it is not only described by the two electromagnetic charges but also by the two gravitoelectromagnetic peers. This provides a remarkable interpretation of the twofold dyonic version of restriction \eqref{rest} as a proportionality relation, mostly via the coupling constants, between a sort of total dyonic electromagnetic charge $q^2+p^2$ and its dyonic gravitational analog $m^2+n^2$, i.e.\ establishing a correlation between both dyonic counterparts.

One can also note that the coupling constants become severely restricted in order to allow the existence of the solution. Indeed, the reality condition on the scalar field \eqref{Phi} will impose that the sign of the cosmological constant $\Lambda$ and the conformal self-interaction coupling constant $\lambda$ must be opposite, while the positiveness of the total dyonic counterparts in relation \eqref{rest} gives rise to the following extra condition
\begin{equation}
\left( k-\frac{4}{3}\Lambda n^2 \right)\!
\!\left(\frac{36}{\kappa}+\frac{\Lambda}{\lambda}\right)>0.
\label{cond}
\end{equation}

In the Maxwell limit $\gamma=0$ the electromagnetic vector potentials become 
\begin{subequations} \label{MaxwellAs}
\begin{align}
A_t(r) &= -\frac{qr}{r^2+n^2} + \frac{p(r^2-n^2)}{2n(r^2+n^2)}, \\
A^{*}_t(r) &= \frac{pr}{r^2+n^2} + \frac{q(r^2-n^2)}{2n(r^2+n^2)},
\end{align}
\end{subequations}
generalizing to the dyonic case the purely electric, $p=0$, conformal Taub-NUT solution of Ref.~\cite{Bardoux:2013swa}. The corresponding metric and scalar field are unaffected by those limits, which brings up the following interesting curiosity of solution \eqref{solution}. On the one hand, the metric and scalar field behave exactly the same way as in the linearly charged case of \cite{Bardoux:2013swa}. On the other hand, the electromagnetic fields of \eqref{solution} are exactly the same as those of ModMax without a scalar field \cite{BallonBordo:2020jtw,Flores-Alfonso:2020nnd}. This peculiarity is a direct consequence of the decoupling between both kinds of involved conformal matter and additionally, has roots in that the constitutive relations \eqref{modmax} turn out to have no information on the particular gravitational potential of the background \eqref{metricAnsatz}. In fact, only the constraint \eqref{rest} nontrivially merges all the involved ingredients. This is in analogy with what occurs in the linear case when the neutral topological MTZ black holes are conventionally charged \cite{Martinez:2002ru,Martinez:2005di}. 

There are other two new nontrivial limits of solution \eqref{solution}, within the list of its underlying properties. The first nontrivial limit is to consider a vanishing cosmological constant, since its naive application in \eqref{solution} could yield a trivial scalar field. In fact, the appropriate way to take this limit is by first solving the constraint \eqref{rest} for the conformal coupling constant $\lambda$ and then evaluating the result in the scalar field \eqref{Phi}. Taking now $\Lambda=0$ will additionally impose $\lambda=0$, and since the resulting metric has only sense for spherical topology $k=1$, this procedure gives rise to the following Taub-NUT generalization of the Bekenstein black hole \cite{Bekenstein:1974sf} but dyonically charged via ModMax electrodynamics 
\begin{subequations}\label{solution_Lambda=0}
\begin{align}
\rd s^2 ={}& -\frac{(r-m)^2}{r^2+n^2}\left(\rd t + 4n\sin^2\tfrac12\theta\rd\phi\right)^2 + \frac{r^2+n^2}{(r-m)^2}\rd r^2\notag\\
& + (r^2+n^2)(\rd\theta^2 + \sin^2\theta\rd\phi^2),
\frac{}{}\\
\Phi ={}& \sqrt{\frac6{\kappa}}
\frac{\sqrt{m^2+n^2
-\frac{\kappa}{8\pi}e^{-\gamma}(q^2+p^2)}}{r-m},\\
A ={}& h(r;p,-q) \left(\rd t +4n\sin^2\tfrac12\theta\rd\phi\right),\frac{}{} \\
A^{*} ={}& h(r;q,p) \left(\rd t +4n\sin^2\tfrac12\theta\rd\phi\right).\frac{}{}
\end{align}
\end{subequations}

The other new nontrivial limit of solution \eqref{solution} is the static one $n\to0$. This limit should be taken with care in the nonlinear electromagnetic sector, since constant contributions of order $1/n$ in the gauge potentials need to be gauged away before taking the limit. After the static limit $n\to0$ is taken with caution we obtain the promised dyonic nonlinearly charged black hole solutions generalizing the MTZ configurations \cite{Martinez:2002ru,Martinez:2005di}
\begin{subequations}\label{solution_n=0}
\begin{align}
\rd s^2 ={}& -\left[-\frac{\Lambda}{3}r^2
+k\left(1-\frac{m}{r}\right)^2\right] \rd t^2\notag\\
& + \left[-\frac{\Lambda}{3}r^2
+k\left(1-\frac{m}{r}\right)^2\right]^{-1}\rd r^2 + r^2\rd\Omega_k^2,\label{metric_n=0}\\
\Phi ={}& \sqrt{-\frac{\Lambda}{6\lambda}}
\frac{m}{r-m},\label{Phi_n=0} \\
A ={}& -\frac{qe^{-\gamma}}{r}\rd t +\frac{2p}{k}\sin^2{(\sqrt{k}\theta/2)}\rd\phi, \\
A^{*} ={}& \frac{pe^{-\gamma}}{r}\rd t +\frac{2q}{k}\sin^2{(\sqrt{k}\theta/2)}\rd\phi,
\end{align}
where the constraint now reflects the loss of dyonic character in the gravitational sector
\begin{equation}
    e^{-\gamma}(q^2+p^2)=\frac{2\pi k}{9}
\!\left(\frac{36}{\kappa}+\frac{\Lambda}{\lambda}\right)\!m^2.
\end{equation}
\end{subequations}
The original charged MTZ black holes \cite{Martinez:2002ru,Martinez:2005di} are recovered in the Maxwell limit $\gamma=0$, and for purely electric configurations $p=0$. Let us stress a crucial difference in the behavior of the solution beyond these further limits. It is clear that in the linear Maxwell case, charged solutions can be purely electric (or magnetic) or even dyonic. In the ModMax case, the constitutive relations \eqref{modmax} become effectively linear when $\sP$ or $\sQ$ vanish. For the black hole \eqref{solution_n=0} we have
\begin{equation}
\sQ=\frac{qpe^{-\gamma}}{r^4},
\end{equation}
and its purely electric (or magnetic) limit would imply $\sQ=0$. Thus, in order to ensure that the new black holes \eqref{solution_n=0} probe the fully nonlinear regime it is mandatory to consider dyonic configurations, as those obtained in Ref.~\cite{Flores-Alfonso:2020euz} in the absence of the scalar field.

As a final comment, when both nontrivial limits are taken simultaneously, i.e.\ $\Lambda=0$, which constrains $\lambda=0$ and $k=1$, together with $n=0$, one gets
\begin{subequations}\label{MMBBekenstein}
\begin{align}
\rd s^2 ={}& -\left(1-\frac{m}{r}\right)^2\rd t^2 + \left(1-\frac{m}{r}\right)^{-2}\rd r^2\notag\\
& + r^2(\rd\theta^2 + \sin^2\theta\rd\phi^2),
\frac{}{}\\
\Phi ={}& \sqrt{\frac{6}{\kappa}}
\frac{\sqrt{m^2-\frac{\kappa}{8\pi}e^{-\gamma}(q^2+p^2)}}{r-m}, \\
A ={}& -\frac{qe^{-\gamma}}{r}\rd t +2p\sin^2{(\theta/2)}\rd\phi, \\
A^{*} ={}& \frac{pe^{-\gamma}}{r}\rd t +2q\sin^2{(\theta/2)}\rd\phi.
\end{align}
\end{subequations}
This constitutes a dyonic ModMax charged generalization of the Bekenstein black hole \cite{Bekenstein:1974sf}.

%%%%%%%%%%%%%%%%%%%%%%%%%%%%%%%%%%%%%%%%%%%%%%%%%%%%%%%
\section{New (super-)renormalizably dressed spacetimes
\label{sec:SR}}
%%%%%%%%%%%%%%%%%%%%%%%%%%%%%%%%%%%%%%%%%%%%%%%%%%%%%%%

In the previous section, we were successful in charging conformally dressed spacetimes beyond the Maxwell theory without spoiling power-counting renormalizability. A further harmless generalization that can be performed now in the scalar sector is to improve its conformal self-interaction potential precisely with power-counting super-renormalizable contributions. The first explicit self-gravitating scalar configurations allowing such contributions were found in Ref.~\cite{Anabalon:2012tu}, and its appearance was intriguing at that moment since it had no precedent at all. It was later explained in \cite{Ayon-Beato:2015ada} how they can be generated from the MTZ black hole \cite{Martinez:2002ru,Martinez:2005di}. The strategy of Ref.~\cite{Ayon-Beato:2015ada} was to devise a generating method for mapping any self-gravitating conformal scalar field with nontrivial self-interaction and cosmological constant to a one-parameter family of a similar system but supplemented with a (super-)renormalizable potential. The devised one-parameter map was the following 
\begin{subequations} \label{transAP}
\begin{align}
\bar{g}_{\mu\nu}&=\left(a\sqrt{\kappa/6}\Phi+1\right)^{2}g_{\mu\nu}, \label{Ctrans}\\
\bar{\Phi}&=\frac{1}{\sqrt{\kappa/6}}\frac{\sqrt{\kappa/6}\Phi+a}{a\sqrt{\kappa/6}\Phi+1}.
\end{align}
The method was also straightforwardly extended to include the Maxwell theory 
\cite{Ayon-Beato:2015ada}, allowing them to find the charged version of the original solutions of \cite{Anabalon:2012tu}, only by scaling the vector potential according to
\begin{equation}
\bar{A}=\sqrt{1-a^2}A.\label{transA}
\end{equation}
Notice that the electromagnetic part of action \eqref{action} is unaffected by \eqref{Ctrans} due to conformal invariance. Hence, as pointed out in Ref.~\cite{Ayon-Beato:2015ada}, for $\gamma=0$ when the Maxwell theory studied there is recovered, the effect of the scaling \eqref{transA} is just to scale the electromagnetic action by the factor $1-a^2$. 

The aim of this section is to generalize the conformally dressed spacetimes \eqref{solution} to new (super-)renormalizably dressed examples by exploiting the results already established in Ref.~\cite{Ayon-Beato:2015ada}. In order to accomplish this goal, we need to extend first the above map to other conformally charged configurations not necessarily rigged by the linearity of Maxwell theory. In particular, to those charged by the ModMax electrodynamics where $\gamma\ne0$ in action \eqref{action}.  Since everything works by scaling the Maxwell action as $1-a^2$, it is enough to demand that under the sought transformation the same occurs for any conformal electromagnetic action. In doing so, we recall that the structural functions of every conformally invariant nonlinear electrodynamics are necessarily homogeneous, as was explained in Sec.~\ref{subsec:CI}. Concretely, let us remember that the degree of the Hamiltonian as a function of the standard invariants must be one; i.e., it satisfies $\sH(\hat{\lambda}\sP,\hat{\lambda}\sQ)=\hat{\lambda}\sH(\sP,\sQ)$. Hence, it is clear that the appropriate transformation for the second vector potential, generalizing the map of Ref.~\cite{Ayon-Beato:2015ada} to include any conformal electrodynamics, must be necessarily given by the same scaling law
\begin{equation}
 \bar{A}^*=\sqrt{1-a^2}A^*. \label{A*}
\end{equation}
\end{subequations}

We are now ready to pursue the strategy of Ref.~\cite{Ayon-Beato:2015ada}. Applying the extended map \eqref{transAP} to ModMax charged configurations transforms the full action \eqref{action} to a new action 
\begin{equation}
\bar{S}[\bar{g},\bar{\Phi},\bar{A},\bar{P}]=(1-a^2)S[g,\Phi,A,P],\label{bSaS}
\end{equation}
defined by
\begin{subequations}\label{barS} {\small 
\begin{equation}
\bar{S}[\bar{g},\bar{\Phi},\bar{A},\bar{P}]=\int \rd^{4}x \sqrt{-\bar{g}}\left(
\frac{\bar{R}-2\bar{\Lambda}}{2\kappa}+\bar{L}_\text{SR}+\bar{L}_\text{MM}\right)\!,
\end{equation}}%
where $\bar{R}$ refers to the scalar curvature associated with the metric $\bar{g}$. Additionally, we have defined $\bar{F}=\rd\bar{A}$ and $\bar{\star}\bar{P}=\rd\bar{A}^*$; accordingly, $\bar{L}_\text{MM}$ is the ModMax Lagrangian \eqref{LNED} for these fields. At the same time, the conformal scalar Lagrangian changes to
\begin{align}
\bar{L}_\text{SR}={}&  - \frac{1}{2}\bar{\nabla}_{\mu}\bar{\Phi}\bar{\nabla}^{\mu}\bar{\Phi} -\frac{1}{12} \bar{R} \bar{\Phi}^2 \notag\\
&-\lambda_1\bar{\Phi} - \lambda_2\bar{\Phi}^2 - \lambda_3\bar{\Phi}^3 - \lambda_4\bar{\Phi}^4,
\end{align}
keeping the conformal coupling to the Ricci scalar, but the corresponding action is no longer conformally invariant due to the new power-counting super-renormalizable contributions improving the self-interaction potential, defined by the emerging coupling constants $\lambda_1$, $\lambda_2$, and $\lambda_3$. In addition, the new cosmological constant is now given by 
\begin{equation}
\bar{\Lambda}=\frac{\kappa\Lambda+36a^{4}\lambda}{\kappa(1-a^{2})^{3}},\label{barLambda}
\end{equation}
while the coupling constants of the potential read
\begin{align}
\lambda_1 &= -\frac{2\sqrt{6}}{3}\frac{a(\kappa\Lambda+36a^{2}\lambda)}
{\kappa^{3/2}(1-a^2)^{3}},\\
\lambda_2 &= \frac{a^{2}(\kappa\Lambda+36\lambda)}{\kappa(1-a^{2})^{3}},\\
\lambda_3 &= -\frac{\sqrt{6}}{9}\frac{a(a^{2}\kappa\Lambda+36\lambda)}
{\kappa^{1/2}(1-a^{2})^{3}},\\
\lambda_4 &= \frac{1}{36}\frac{a^{4}\kappa\Lambda+36\lambda}{(1-a^{2})^{3}},
\label{lambda4}
\end{align}
\end{subequations}
and all are parameterized in terms of the couplings of the starting theory together with the parameter $a$. In summary, self-gravitating ModMax charged solutions of a conformal scalar field generate by means of the transformation \eqref{transAP} a one-parameter family of also self-gravitating ModMax charged solutions of a conformally coupled scalar field but now ruled by a (super-)renormalizable self-interaction.

Taking the solution \eqref{solution} as a seed configuration for the transformation \eqref{transAP}, we are now able to generate a (super-)renormalizably dressed solution that is an extreme of action \eqref{barS}, and is given by 
\begin{subequations} \label{superrernom}
\begin{align}
\rd\bar{s}^2 ={}& \left(\frac{r-m+a\sqrt{-\frac{\kappa\Lambda}{36\lambda}(m^2+n^2)}}{r-m}\right)^2 \notag\\ 
 & \times \Biggr[ -f(r)\left(\rd t +\frac{4n}k\sin^2{(\sqrt{k}\theta/2)}\rd\phi\right)^2
\notag \\
& \quad\,\,+ \frac{\rd r^2}{f(r)} + (r^2+n^2)\rd\Omega_k^2\,\Biggr],\\
\bar{\Phi} ={}& \sqrt{\frac{6}{\kappa}}\,
\frac{a(r-m)+\sqrt{-\frac{\kappa\Lambda}{36\lambda}(m^2+n^2)}}
{r-m+a\sqrt{-\frac{\kappa\Lambda}{36\lambda}(m^2+n^2)}},\\
\bar{A} ={}& h(r;\bar{p},-\bar{q})\left(\rd t +\frac{4n}k\sin^2{(\sqrt{k}\theta/2)}\rd\phi\right),
 \label{barA}\\
\bar{A}^{*} ={}& h(r;\bar{q},\bar{p}) \left(\rd t +\frac{4n}k\sin^2{(\sqrt{k}\theta/2)}\rd\phi\right),
 \label{barA*}
\end{align}
where the gravitational and vector potentials are again determined by the functions \eqref{metricFunction} and \eqref{h}, respectively. The latter function is now parameterized by the new constants $\bar{q}=\sqrt{1-a^2}q$ and $\bar{p}=\sqrt{1-a^2}p$, which are subject to the constraint 
\begin{align}
e^{-\gamma}(\bar{q}^2+\bar{p}^2) ={}&(1-a^2)\frac{2\pi}{9}
\left( k-\frac{4}{3}\Lambda n^2 \right)
\notag\\
&\times\left(\frac{36}{\kappa}+\frac{\Lambda}{\lambda}\right)
(m^2+n^2).
\end{align}
\end{subequations}
These new constants are correspondingly the electromagnetic charges in this context, as can be straightforwardly checked in the spherical case $k=1$ where 
\begin{equation}
\bar{p}=\frac{1}{4\pi}\int_{\partial\Sigma}\bar{F},\qquad 
\bar{q}=\frac{1}{4\pi}\int_{\partial\Sigma}\bar{\star}\bar{P},
\end{equation}
with $\partial\Sigma$ consequently taken as a sphere at infinity. It is worth mentioning that for purely electric configurations, $p=0$, of the Maxwell case, $\gamma=0$, the above solution becomes that of Ref.~\cite{Barrientos:2022avi}. Consequently, setting $a=0$ in the (super-)renormalizably dressed solution \eqref{superrernom} one recovers the conformally dressed seed \eqref{solution}. The other way around, taking as seeds all the special limits of \eqref{solution}---namely \eqref{solution_Lambda=0}, \eqref{solution_n=0}, and \eqref{MMBBekenstein}---new special (super-)renormalizably dressed solutions are straightforwardly obtained; since those limits commute with the map there is no need to explicitly write them.

%%%%%%%%%%%%%%%%%%%%%%%%%%%%%%%%%%%%%%%%%%%%%%%%%%%%%%%%%%%%%%%%%%%%%%%%%%%%%%%%%%%%%%%%%
\section{Nonlinearly charging the non-Noetherian conformal sector\label{sec:nonNoether}}
%%%%%%%%%%%%%%%%%%%%%%%%%%%%%%%%%%%%%%%%%%%%%%%%%%%%%%%%%%%%%%%%%%%%%%%%%%%%%%%%%%%%%%%%%

It is certainly known now that the most general second-order equation describing a conformally invariant scalar field that is the extreme of an action principle is definitely not Eq.~\eqref{Eq_CS}, as it has been recently proved in \cite{Ayon-Beato:2023bzp} inspired in the previous work of \cite{Fernandes:2021dsb}. In particular, it was originally identified in Ref.~\cite{Fernandes:2021dsb} that a non-Noetherian conformal contribution can be also considered in Eq.~\eqref{Eq_CS}. It comes from the following additional piece to Lagrangian \eqref{L_CS} that extends the action principle \eqref{action} used in Sec.~\ref{sec:MMchargedMTZ} to
\begin{subequations}\label{actionNN}
\begin{equation}
S[g,\Phi,A,P]+\int d^4x\sqrt{-g}L_\text{NN},
\end{equation}
with a non-Noetherian conformal Lagrangian density defined by {\small
\begin{align}
L_\text{NN}=&-\frac\alpha2
\biggl(\ln(\Phi)\mathscr{G} 
-\frac4{\Phi^2}G^{\mu\nu}\nabla_\mu\Phi\nabla_\nu\Phi
-\frac4{\Phi^3}(\nabla_\mu\Phi\nabla^\mu\Phi)\Box\Phi
\notag \\ 
&\qquad\
+ \frac2{\Phi^4}(\nabla_\mu\Phi\nabla^\mu\Phi)^2\biggr),
\label{LnonNoether}
\end{align}}%
\end{subequations}
where $\mathscr{G}=R^2-4R_{\alpha\beta}R^{\alpha\beta}
+R_{\alpha\beta\mu\nu}R^{\alpha\beta\mu\nu}$ is the Gauss-Bonnet density, $G^{\mu\nu}$ the Einstein tensor, and $\Box$ the d'Alembert operator, all being defined with the standard metric $g_{\mu\nu}$. With this new contribution, Eq.~\eqref{Eq_CS} is extended as
\begin{equation}
\Box\Phi-\frac16R\Phi-\Phi^3\left(4\lambda+\frac\alpha2\tilde{\mathscr{G}}\right)=0,
\label{EqnonNoether}
\end{equation}
where $\tilde{\mathscr{G}}$ is the Gauss-Bonnet density again, but now corresponding to the auxiliary metric $\tilde{g}_{\mu\nu}=\Phi^2g_{\mu\nu}$, which by construction is invariant under conformal transformations 
\begin{equation}
(g_{\mu\nu},\Phi)\,\mapsto\,
(\Omega^2g_{\mu\nu},\Omega^{-1}\Phi).
\label{CTPhi}
\end{equation}
Since the new term with the tilde in \eqref{EqnonNoether} behaves as a constant under conformal transformations, the resulting equation is again conformally invariant. However, the action built from Lagrangian \eqref{LnonNoether} is not preserved under the transformations \eqref{CTPhi} and consequently, the conformal invariance of \eqref{EqnonNoether} is not due to the Noether theorem. This is why Ref.~\cite{Ayon-Beato:2023bzp} termed non-Noetherian conformal scalar fields to those whose dynamic is governed by the previous equation.\footnote{For the two-dimensional non-Noetherian analog see Refs.~\cite{Jackiw:2005su,Ayon-Beato:2023lrn}.} 

As an Euler-Lagrange equation of a related action, the second-order conformally invariant properties of \eqref{EqnonNoether} are not spoiled by adding to Lagrangian \eqref{LnonNoether} arbitrary nonminimal couplings to gravity build with the Weyl tensor of the auxiliary metric $\tilde{g}_{\mu\nu}$ \cite{Ayon-Beato:2023bzp}. Nevertheless, only the extension \eqref{LnonNoether} also gives rise to a second-order addition to the energy-momentum tensor \eqref{Tmunu}, since the Lagrangian \eqref{LnonNoether} remarkably belongs to the most general family of second-order variational principles for the scalar field and the metric which are defined by the Horndeski action \cite{Horndeski:1974wa}. We do not write here the involved explicit expression of the new contribution to the energy-momentum tensor for the economy of space, but it can be consulted in Ref.~\cite{Fernandes:2021dsb}.

The static spherically symmetric non-Noetherian conformally dressed black holes, charged with the $\gamma=0$ Maxwell sector of action \eqref{actionNN}, were originally derived in Ref.~\cite{Fernandes:2021dsb}, where Fernandes showed the existence of two independent branches of solutions. There is no known Taub-NUT generalization of the Fernandes static branches. Hence, here we focus on the solutions of the non-Noetherianly extended conformal system derived from action \eqref{actionNN} in the static limit, $n=0$, of our study ansatz \eqref{Taub-NUTansatz}. When the ModMax electrodynamics is used to nonlinearly charge the two Fernandes branches of non-Noetherian conformal black holes beyond the Maxwell regime, such branches are generalized as follows {\small
\begin{subequations}\label{nonNoetherSol} 
\begin{align}
\rd s^2 ={}& - 
\Biggl\{ k+\frac{r^2}{2\kappa\alpha}
\Biggl[ 1 \pm \sqrt{1+4\kappa\alpha
\left(\frac{\Lambda}3+\frac{2m(r)}{r^3}\right)}
\Biggl] \Biggr\} \rd t^2 \notag \\
&+ 
\Biggl\{ k+\frac{r^2}{2\kappa\alpha}
\Biggl[ 1 \pm \sqrt{1+4\kappa\alpha
\left(\frac{\Lambda}3+\frac{2m(r)}{r^3}\right)}
\Biggl] \Biggr\}^{-1}\rd r^2 \notag \\
&+ r^2\rd\Omega_k^2,\frac{}{}\label{gFer}\\
A ={}& -\frac{qe^{-\gamma}}{r}\rd t +\frac{2p}{k}\sin^2{(\sqrt{k}\theta/2)}\rd\phi, \\
A^{*} ={}& \frac{pe^{-\gamma}}{r}\rd t +\frac{2q}{k}\sin^2{(\sqrt{k}\theta/2)}\rd\phi,
\end{align}
\end{subequations}}%
where the metric function and the scalar field are given in one case as
\begin{subequations}
\begin{align}
m(r)&=m-\frac{\kappa}{16\pi}
\frac{e^{-\gamma}(q^2+p^2)-16\pi k^2\alpha}r,\\
\Phi(r) &=\frac{\sqrt{-12k\alpha}}r, \\
\lambda\alpha&=\frac{1}{144},
\end{align}
\end{subequations}
while the second branch depends on the sign of the non-Noetherian conformal coupling constant $\alpha$ as
\begin{subequations}
\begin{align}
m(r)&=m-\frac{\kappa}{16\pi}
\frac{e^{-\gamma}(q^2+p^2)}r,\\
\Phi(r)&=
\begin{cases}
\dfrac{\sqrt{12k\alpha}}{r\sinh\left(\sqrt{k}\left[c \pm {\displaystyle\int}\!\dfrac{\rd r}{r\sqrt{f}}\right]\right)}, & \alpha>0, \\
&\\
\dfrac{\sqrt{-12k\alpha}}{r\cosh\left(\sqrt{k}\left[c \pm {\displaystyle\int}\!\dfrac{\rd r}{r\sqrt{f}}\right]\right)}, & \alpha<0,
\end{cases}\label{PhiFerb}\\
\lambda\alpha&=\frac1{48}.
\end{align}
\end{subequations}
Here $c$ is a sort of non-Noetherian conformal hair and the signs in the scalar field \eqref{PhiFerb} are independent of those of the metric \eqref{gFer}.

For $\gamma=0$ and $k=1$ the conventionally charged spherically symmetric Fernandes black holes \cite{Fernandes:2021dsb}, dressed with a non-Noetherian conformal scalar field, are recovered. In order to ease the comparison we make transparent the relation between the Fernandes notations and ours
\begin{align}
\beta^\text{F}  &=\frac{\kappa}6, &
\lambda^\text{F}&=\kappa\lambda, &
\alpha^\text{F} &=\kappa\alpha, \notag\\
A^\text{F}_\mu  &=\frac1{\sqrt{4\pi}}A_\mu, &
Q^\text{F}_e    &=\sqrt{4\pi}q, &
Q^\text{F}_m    &=\sqrt{4\pi}p.
\end{align}

We found it appropriate to emphasize the further bifurcation \eqref{PhiFerb} of the second branch not reported in Ref.~\cite{Fernandes:2021dsb}, since the spherical $\alpha<0$ sector originally found in that reference is not extensible to flat $k=0$ or hyperbolic $k=-1$ topologies, which are only covered by the new $\alpha>0$ sector. In fact, the latter is the only non-Noetherian conformally dressed solution at all surviving in the limit $k\to0$. This is compatible with the recent results of Refs.~\cite{Babichev:2023rhn} and \cite{Hassaine:2023paj} for more general, non-necessarily symmetric, Horndeski theories allowing a similar branching structure to the one first found by Fernandes in Ref.~\cite{Fernandes:2021dsb}. There, generic planar solutions $k=0$ unavoidably require the absence of the standard Noetherian conformal sector $L_\text{CS}$ \eqref{L_CS}, which is impossible here. Besides, if the limit $\alpha=0$ is assumed then the metric lower-sign solution becomes the ModMax nonlinearly charged black holes of \cite{Flores-Alfonso:2020euz} generalizing the dyonic Reissner-Nordstr\"om ones, where the scalar field is absent.

Finally, the fact that the matter part of the new action \eqref{actionNN} is no longer conformally invariant makes it impossible to straightforwardly apply the methods of Ref.~\cite{Ayon-Beato:2015ada} to explore (super-)renormalizably dressed configurations. How such a strategy can be extended to include the new non-Noetherian conformal sector remains an open problem for the future.

%%%%%%%%%%%%%%%%%%%%%%%%%%%%%%%%%%%%%%%%%%%%%%
\section{Concluding Remarks\label{sec:Conclu}}
%%%%%%%%%%%%%%%%%%%%%%%%%%%%%%%%%%%%%%%%%%%%%%

In this work, we have considered self-gravitating configurations of ModMax nonlinear electrodynamics together with a self-interacting scalar field conformally coupled to gravity. The distinctive feature of ModMax theory is that it is the unique nonlinear extension of the Maxwell theory enjoying its conformal invariance as well as its duality symmetry \cite{Bandos:2020jsw}. We have taken advantage of this work to present a novel and transparent derivation of this electrodynamics by employing the dual Legendre formulations introduced in Ref.~\cite{Salazar:1987ap}. Of particular relevance is their simple characterization of duality-invariant theories in contrast with that allowed for the standard Lagrangian formalism \cite{Russo:2024llm}. Our approach demands that the Legendre duals themselves be invariant under duality and conformal transformations. In doing so, we end up with elementary linear first-order equations which straightforwardly yield as unique solution the ModMax electrodynamics \cite{Bandos:2020jsw}. 

Correspondingly, a limitation of our work is the same as the approach that we borrow from Ref.~\cite{Salazar:1987ap} to duality invariance, i.e.\ it only applies to the most common algebraically general electromagnetic configurations where not all the eigenvalues vanish \cite{Plebanski:1970}, making sense of the alignment \eqref{aligned}. This means that algebraically special or null electromagnetic fields, where all the eigenvalues are identically zero \cite{Plebanski:1970}, are not included in our analysis. Therefore, the use of different variables including the particular case of null fields can still give rise to potentially different realizations of duality invariance excluded from those treated here. This seems to be the case of those configurations rigged by the Bia{\l}ynicki-Birula electrodynamics that is invariant under $SL(2,\mathbb{R})$ transformations more than duality rotations and additionally also respects conformal symmetry \cite{Bialynicki-Birula:1984daz,Bialynicki-Birula:1992rcm}.

Regarding the ModMax nonlinearly charged spacetimes, our most general result is the family of solutions \eqref{superrernom}, which describes nonstatic configurations whose background is dyonically charged by ModMax fields and is simultaneously dressed by a conformally coupled and (super-)renormalizable scalar field. Algebraically, the spacetime is classified in the Petrov scheme as of type $D$. Whereas, geometrically, it is conformally related to a complex line bundle fibered on a K\"ahler manifold. As such, the solution is closely related to the classical Taub-NUT vacuum metric \cite{Taub:1950ez,Newman:1963yy,Misner:1963fr}. In the static limit, the fibration is trivialized and also black hole solutions emerge. In that case, the resulting black holes are nonlinearly charged generalizations of those presented in Ref.~\cite{Ayon-Beato:2015ada}, which in turn has as neutral limits the ones found in Ref.~\cite{Anabalon:2012tu}.

In the more general stationary setup, many special spacetimes were shown to exist in \cite{Barrientos:2022avi} for the purely electric Maxwell case such as wormholes, cosmological bounces, and regular black holes. The same configurations also appear here since our metric is formally indistinguishable from theirs, except that it is now supported by a more general content composed of dyonic nonlinearly charged matter. Since such metrics were thoroughly studied in Ref.~\cite{Barrientos:2022avi} there is no need to repeat their analysis here.

The obtained (super-)renormalizably dressed spacetimes are possible because the four coupling constants characterizing the scalar field self-interaction potential are not all independent. Indeed, only a one-parameter subspace of the emerging strictly super-renormalizable sector is available to the configuration. This is a direct consequence of the solution-generating method proposed in Ref.~\cite{Ayon-Beato:2015ada} and generalized here to any conformal electrodynamics, that maps a self-gravitating conformal seed into the larger parameter space of the self-gravitating (super-)renormalizable theories. 

In order to cover the whole spectrum of conformally dressed black holes, we have also addressed the extension to the ModMax regime of the recently discussed non-Noetherian conformal scalar fields \cite{Fernandes:2021dsb,Ayon-Beato:2023bzp}. We were able to nonlinearly charge the non-Noetherian conformally dressed black holes of Ref.~\cite{Fernandes:2021dsb}. Instead, we leave open the problem of generating (super-)renormalizably dressed spacetimes in this context due to the impossibility of using the generating methods of Ref.~\cite{Ayon-Beato:2015ada} by the non-Noetherian origin of conformal symmetry in the scalar fields.

We expect the emergence of all the presented solutions definitely encourages further research into the fertile subject of scalar fields conformally coupled to gravity and supplemented by nonlinear electrodynamics, especially in light of the recent developments discussed in Refs.~\cite{Barrientos:2022yoz,Bravo-Gaete:2022mnr,Anastasiou:2022wjq,Colipi-Marchant:2023awk}. A very promising path within all the potential possibilities are new results \cite{Podolsky:2021zwr,Podolsky:2022xxd,Astorino:2023elf,Astorino:2023uim,Astorino:2024bfl} on the celebrated Pleba\'nski-Demia\'nski solutions \cite{Plebanski:1976gy}, since thanks to them the presence of acceleration and NUT parameter in Petrov type D spacetimes is now better understood. This finally clarifies the puzzle of why the accelerated NUT vacuum spacetimes originally found in \cite{Chng:2006gh} are outside this special class, being in fact of general Petrov type \cite{Podolsky:2020xkf}. The connection with our work is that on the one hand, the accelerated NUT spacetimes have been recently extended to be sourced by free conformal scalar fields \cite{Barrientos:2023tqb}, and on the other hand, the first nonlinearly charged accelerated spacetimes were precisely built with ModMax \cite{Barrientos:2022bzm}.

%%%%%%%%%%%%%%%%%%%%%%%%%%%%%%%

\begin{acknowledgments}
The authors would like to thank the NordGrav Summer Workshop recently held at the Instituto de Ciencias Exactas y Naturales, Universidad Arturo Prat, partially supported by the Agencia Nacional de Investigaci\'on y Desarrollo (ANID) under FONDECYT Grant No.~11200742. D.F.-A.\ would like to thank the Departamento de F\'{i}sica of CINVESTAV-IPN for their kind hospitality and stimulating environment. We thank M.\ Astorino, L.\ Avil\'es, J.\ Barrientos, A.\ Cisterna, C.\ Corral, P.\ Fernandes, J.\ Oliva, and P.A.\ S\'anchez for interesting discussions and suggestions. E.A.-B.\ has been partially funded by Conahcyt Grant No.~A1-S-11548. D.F.-A.\ is supported by ANID under FONDECYT Grant No.~3220083. M.H.\ has been partially funded by FONDECYT Grant No.~1210889.
\end{acknowledgments}

\appendix*

\section{General duality-invariant Lagrangians are necessarily implicit}

Using the methods recently revised in Ref.~\cite{Ayon-Beato:2024xgp} it is possible to show that, for example, the general solution to the first of the nonlinear first-order PDE \eqref{nl1oPDE} determining all duality-invariant Lagrangians is
\begin{subequations}\label{L_D(E,B)}
\begin{equation}
\sL(E,B)=-\frac12\left(e^{\Gamma}E^2
-e^{-\Gamma}B^2\right)+J(\Gamma),
\end{equation}
where $\Gamma(E,B)$ is a function determined by the arbitrary implicit dependence
\begin{equation}
J'(\Gamma)=\frac12\left(e^{\Gamma}E^2
+e^{-\Gamma}B^2\right).\label{constrEB}
\end{equation}
\end{subequations}
It is straightforward to show that \eqref{L_D(E,B)} is a solution to the first of PDE \eqref{nl1oPDE} if and only if the following conditions are satisfied 
\begin{subequations}\label{Rconstr}
\begin{align}
\left[J'(\Gamma)-\frac12\left(e^{\Gamma}E^2
+e^{-\Gamma}B^2\right)\right]\Gamma_E&=0,\\
\left[J'(\Gamma)-\frac12\left(e^{\Gamma}E^2
+e^{-\Gamma}B^2\right)\right]\Gamma_B&=0,
\end{align}
\end{subequations}
which are warranted by \eqref{constrEB}. Since this solution involves the single-argument arbitrary function $J$, it is in fact the general solution. Such implicit dependence generically characterizes all duality-invariant nonlinear electrodynamics in the Lagrangian formulation; see the recent results of Ref.~\cite{Russo:2024llm} for a similar representation in different variables that we revise in a moment. It is important to emphasize that, as conditions \eqref{Rconstr} clearly indicate, the implicit dependence \eqref{constrEB} is strictly needed to have solutions only when the function $\Gamma$ is nontrivial. In the degenerate case where such function is a constant, \eqref{constrEB} is unnecessary and solution \eqref{L_D(E,B)} becomes the standard sum separable solution. This degenerate sum-separable case is precisely the one corresponding to ModMax
\begin{equation}
\Gamma_\text{MM}=\gamma=\text{const.}\frac{}{}\label{GaMMa}
\end{equation}
where in all generality we can choose $J=0$ in \eqref{L_D(E,B)}, to avoid redefining the cosmological constant, recovering the ModMax Lagrangian \eqref{L_MM(E,B)}. In terms of the standard invariants the Lagrangian characterizing all duality-invariant theories is written as follows {\small 
\begin{subequations}\label{dualityL}
\begin{align}
\sL(\sF,\sG)&=\cosh(\Gamma)\sF
-\sinh(\Gamma)\sqrt{\sF^2+\sG^2}+J(\Gamma),\\
0&=\sinh(\Gamma)\sF
-\cosh(\Gamma)\sqrt{\sF^2+\sG^2}+J'\!(\Gamma),
\label{constrFG}
\end{align}
\end{subequations}}%
where the last relation implicitly determines the function $\Gamma(\sF,\sG)$ for each theory defined by the single argument function $J$. The only exception is the ModMax theory \eqref{GaMMa}, as has already been explained. 

ModMax is characterized differently in Ref.~\cite{Russo:2024llm}, where their implicit representation for general duality-invariant Lagrangians is based instead on the product separable solutions giving {\small
\begin{equation}
\sL(E,B)=B\sqrt{2\tau-E^2}-v(\tau),\quad
\dot{v}(\tau)=\frac{B}{\sqrt{2\tau-E^2}}.\label{L_D(E,B)_RT}
\end{equation}}%
As was emphasized in \cite{Russo:2024llm}, within this context ModMax is recovered by choosing the function 
\begin{equation}
v_\text{MM}(\tau)=e^{\gamma}\tau.
\end{equation}
We give the explicit relation between the notations of Russo and Townsend in \cite{Russo:2024llm} with ours to facilitate the comparison {\small
\begin{align}
S^\text{RT}&=-\sF, & P^\text{RT}&=-\sG, & \mathcal{L}^\text{RT}&=-\sL, \notag\\ 
U^\text{RT}&=\frac{B^2}2, & V^\text{RT}&=\frac{E^2}2. &&
\end{align}}%
Interestingly, both representations \eqref{L_D(E,B)} and \eqref{L_D(E,B)_RT}
are related by a Legendre transform between the arbitrary functions
\begin{equation}
J(\Gamma)=e^{\Gamma}\tau-v(\tau), \quad e^{\Gamma}=\dot{v}.
\end{equation}
Notice that it becomes degenerate precisely in the ModMax case.


\begin{thebibliography}{71}%
\makeatletter
\providecommand \@ifxundefined [1]{%
 \@ifx{#1\undefined}
}%
\providecommand \@ifnum [1]{%
 \ifnum #1\expandafter \@firstoftwo
 \else \expandafter \@secondoftwo
 \fi
}%
\providecommand \@ifx [1]{%
 \ifx #1\expandafter \@firstoftwo
 \else \expandafter \@secondoftwo
 \fi
}%
\providecommand \natexlab [1]{#1}%
\providecommand \enquote  [1]{``#1''}%
\providecommand \bibnamefont  [1]{#1}%
\providecommand \bibfnamefont [1]{#1}%
\providecommand \citenamefont [1]{#1}%
\providecommand \href@noop [0]{\@secondoftwo}%
\providecommand \href [0]{\begingroup \@sanitize@url \@href}%
\providecommand \@href[1]{\@@startlink{#1}\@@href}%
\providecommand \@@href[1]{\endgroup#1\@@endlink}%
\providecommand \@sanitize@url [0]{\catcode `\\12\catcode `\$12\catcode
  `\&12\catcode `\#12\catcode `\^12\catcode `\_12\catcode `\%12\relax}%
\providecommand \@@startlink[1]{}%
\providecommand \@@endlink[0]{}%
\providecommand \url  [0]{\begingroup\@sanitize@url \@url }%
\providecommand \@url [1]{\endgroup\@href {#1}{\urlprefix }}%
\providecommand \urlprefix  [0]{URL }%
\providecommand \Eprint [0]{\href }%
\providecommand \doibase [0]{https://doi.org/}%
\providecommand \selectlanguage [0]{\@gobble}%
\providecommand \bibinfo  [0]{\@secondoftwo}%
\providecommand \bibfield  [0]{\@secondoftwo}%
\providecommand \translation [1]{[#1]}%
\providecommand \BibitemOpen [0]{}%
\providecommand \bibitemStop [0]{}%
\providecommand \bibitemNoStop [0]{.\EOS\space}%
\providecommand \EOS [0]{\spacefactor3000\relax}%
\providecommand \BibitemShut  [1]{\csname bibitem#1\endcsname}%
\let\auto@bib@innerbib\@empty
%</preamble>
\bibitem [{\citenamefont {Heisenberg}\ and\ \citenamefont
  {Euler}(1936)}]{Heisenberg:1936nmg}%
  \BibitemOpen
  \bibfield  {author} {\bibinfo {author} {\bibfnamefont {W.}~\bibnamefont
  {Heisenberg}}\ and\ \bibinfo {author} {\bibfnamefont {H.}~\bibnamefont
  {Euler}},\ }\href {https://doi.org/10.1007/BF01343663} {\bibfield  {journal}
  {\bibinfo  {journal} {Z. Phys.}\ }\textbf {\bibinfo {volume} {98}},\ \bibinfo
  {pages} {714} (\bibinfo {year} {1936})},\ \Eprint
  {https://arxiv.org/abs/physics/0605038} {arXiv:physics/0605038} \BibitemShut
  {NoStop}%
\bibitem [{\citenamefont {Weisskopf}(1936)}]{Weisskopf:406571}%
  \BibitemOpen
  \bibfield  {author} {\bibinfo {author} {\bibfnamefont {V.~F.}\ \bibnamefont
  {Weisskopf}},\ }\href {https://cds.cern.ch/record/406571} {\bibfield
  {journal} {\bibinfo  {journal} {Dan. Mat. Fys. Medd.}\ }\textbf {\bibinfo
  {volume} {14}},\ \bibinfo {pages} {1} (\bibinfo {year} {1936})}\BibitemShut
  {NoStop}%
\bibitem [{\citenamefont {Dunne}(2004)}]{Dunne:2004nc}%
  \BibitemOpen
  \bibfield  {author} {\bibinfo {author} {\bibfnamefont {G.~V.}\ \bibnamefont
  {Dunne}},\ }\bibinfo {title} {{Heisenberg-Euler effective Lagrangians: Basics
  and extensions}},\ in\ \href {https://doi.org/10.1142/9789812775344_0014}
  {\emph {\bibinfo {booktitle} {{From fields to strings: Circumnavigating
  theoretical physics}}}}\ (\bibinfo  {publisher} {World Scientific},\ \bibinfo
  {address} {Singapore},\ \bibinfo {year} {2004})\ pp.\ \bibinfo {pages}
  {445--522},\ \Eprint {https://arxiv.org/abs/hep-th/0406216}
  {arXiv:hep-th/0406216} \BibitemShut {NoStop}%
\bibitem [{\citenamefont {Ellis}\ \emph {et~al.}(2022)\citenamefont {Ellis},
  \citenamefont {Mavromatos}, \citenamefont {Roloff},\ and\ \citenamefont
  {You}}]{Ellis:2022uxv}%
  \BibitemOpen
  \bibfield  {author} {\bibinfo {author} {\bibfnamefont {J.}~\bibnamefont
  {Ellis}}, \bibinfo {author} {\bibfnamefont {N.~E.}\ \bibnamefont
  {Mavromatos}}, \bibinfo {author} {\bibfnamefont {P.}~\bibnamefont {Roloff}},\
  and\ \bibinfo {author} {\bibfnamefont {T.}~\bibnamefont {You}},\ }\href
  {https://doi.org/10.1140/epjc/s10052-022-10565-w} {\bibfield  {journal}
  {\bibinfo  {journal} {Eur. Phys. J. C}\ }\textbf {\bibinfo {volume} {82}},\
  \bibinfo {pages} {634} (\bibinfo {year} {2022})},\ \Eprint
  {https://arxiv.org/abs/2203.17111} {arXiv:2203.17111 [hep-ph]} \BibitemShut
  {NoStop}%
\bibitem [{\citenamefont {Born}\ and\ \citenamefont
  {Infeld}(1934)}]{Born:1934gh}%
  \BibitemOpen
  \bibfield  {author} {\bibinfo {author} {\bibfnamefont {M.}~\bibnamefont
  {Born}}\ and\ \bibinfo {author} {\bibfnamefont {L.}~\bibnamefont {Infeld}},\
  }\href {https://doi.org/10.1098/rspa.1934.0059} {\bibfield  {journal}
  {\bibinfo  {journal} {Proc. Roy. Soc. Lond. A}\ }\textbf {\bibinfo {volume}
  {144}},\ \bibinfo {pages} {425} (\bibinfo {year} {1934})}\BibitemShut
  {NoStop}%
\bibitem [{\citenamefont {Fradkin}\ and\ \citenamefont
  {Tseytlin}(1985)}]{Fradkin:1985qd}%
  \BibitemOpen
  \bibfield  {author} {\bibinfo {author} {\bibfnamefont {E.~S.}\ \bibnamefont
  {Fradkin}}\ and\ \bibinfo {author} {\bibfnamefont {A.~A.}\ \bibnamefont
  {Tseytlin}},\ }\href {https://doi.org/10.1016/0370-2693(85)90205-9}
  {\bibfield  {journal} {\bibinfo  {journal} {Phys. Lett. B}\ }\textbf
  {\bibinfo {volume} {163}},\ \bibinfo {pages} {123} (\bibinfo {year}
  {1985})}\BibitemShut {NoStop}%
\bibitem [{\citenamefont {Bachas}(1996)}]{Bachas:1995kx}%
  \BibitemOpen
  \bibfield  {author} {\bibinfo {author} {\bibfnamefont {C.}~\bibnamefont
  {Bachas}},\ }\href {https://doi.org/10.1016/0370-2693(96)00238-9} {\bibfield
  {journal} {\bibinfo  {journal} {Phys. Lett. B}\ }\textbf {\bibinfo {volume}
  {374}},\ \bibinfo {pages} {37} (\bibinfo {year} {1996})},\ \Eprint
  {https://arxiv.org/abs/hep-th/9511043} {arXiv:hep-th/9511043} \BibitemShut
  {NoStop}%
\bibitem [{\citenamefont {Ay{\'{o}}n-Beato}\ and\ \citenamefont
  {Garc{\'{i}}a}(1998)}]{Ayon-Beato:1998hmi}%
  \BibitemOpen
  \bibfield  {author} {\bibinfo {author} {\bibfnamefont {E.}~\bibnamefont
  {Ay{\'{o}}n-Beato}}\ and\ \bibinfo {author} {\bibfnamefont {A.}~\bibnamefont
  {Garc{\'{i}}a}},\ }\href {https://doi.org/10.1103/PhysRevLett.80.5056}
  {\bibfield  {journal} {\bibinfo  {journal} {Phys. Rev. Lett.}\ }\textbf
  {\bibinfo {volume} {80}},\ \bibinfo {pages} {5056} (\bibinfo {year}
  {1998})},\ \Eprint {https://arxiv.org/abs/gr-qc/9911046}
  {arXiv:gr-qc/9911046} \BibitemShut {NoStop}%
\bibitem [{\citenamefont {Ay{\'{o}}n-Beato}\ and\ \citenamefont
  {Garc{\'{i}}a}(1999{\natexlab{a}})}]{Ayon-Beato:1999qin}%
  \BibitemOpen
  \bibfield  {author} {\bibinfo {author} {\bibfnamefont {E.}~\bibnamefont
  {Ay{\'{o}}n-Beato}}\ and\ \bibinfo {author} {\bibfnamefont {A.}~\bibnamefont
  {Garc{\'{i}}a}},\ }\href {https://doi.org/10.1023/A:1026640911319} {\bibfield
   {journal} {\bibinfo  {journal} {Gen. Rel. Grav.}\ }\textbf {\bibinfo
  {volume} {31}},\ \bibinfo {pages} {629} (\bibinfo {year}
  {1999}{\natexlab{a}})},\ \Eprint {https://arxiv.org/abs/gr-qc/9911084}
  {arXiv:gr-qc/9911084} \BibitemShut {NoStop}%
\bibitem [{\citenamefont {Ay{\'{o}}n-Beato}\ and\ \citenamefont
  {Garc{\'{i}}a}(1999{\natexlab{b}})}]{Ayon-Beato:1999kuh}%
  \BibitemOpen
  \bibfield  {author} {\bibinfo {author} {\bibfnamefont {E.}~\bibnamefont
  {Ay{\'{o}}n-Beato}}\ and\ \bibinfo {author} {\bibfnamefont {A.}~\bibnamefont
  {Garc{\'{i}}a}},\ }\href {https://doi.org/10.1016/S0370-2693(99)01038-2}
  {\bibfield  {journal} {\bibinfo  {journal} {Phys. Lett. B}\ }\textbf
  {\bibinfo {volume} {464}},\ \bibinfo {pages} {25} (\bibinfo {year}
  {1999}{\natexlab{b}})},\ \Eprint {https://arxiv.org/abs/hep-th/9911174}
  {arXiv:hep-th/9911174} \BibitemShut {NoStop}%
\bibitem [{\citenamefont {Ay{\'{o}}n-Beato}\ and\ \citenamefont
  {Garc{\'{i}}a}(2005)}]{Ayon-Beato:2004ywd}%
  \BibitemOpen
  \bibfield  {author} {\bibinfo {author} {\bibfnamefont {E.}~\bibnamefont
  {Ay{\'{o}}n-Beato}}\ and\ \bibinfo {author} {\bibfnamefont {A.}~\bibnamefont
  {Garc{\'{i}}a}},\ }\href {https://doi.org/10.1007/s10714-005-0050-y}
  {\bibfield  {journal} {\bibinfo  {journal} {Gen. Rel. Grav.}\ }\textbf
  {\bibinfo {volume} {37}},\ \bibinfo {pages} {635} (\bibinfo {year} {2005})},\
  \Eprint {https://arxiv.org/abs/hep-th/0403229} {arXiv:hep-th/0403229}
  \BibitemShut {NoStop}%
\bibitem [{\citenamefont {Vagnozzi}\ \emph {et~al.}(2023)\citenamefont
  {Vagnozzi} \emph {et~al.}}]{Vagnozzi:2022moj}%
  \BibitemOpen
  \bibfield  {author} {\bibinfo {author} {\bibfnamefont {S.}~\bibnamefont
  {Vagnozzi}} \emph {et~al.},\ }\href
  {https://doi.org/10.1088/1361-6382/acd97b} {\bibfield  {journal} {\bibinfo
  {journal} {Class. Quant. Grav.}\ }\textbf {\bibinfo {volume} {40}},\ \bibinfo
  {pages} {165007} (\bibinfo {year} {2023})},\ \Eprint
  {https://arxiv.org/abs/2205.07787} {arXiv:2205.07787 [gr-qc]} \BibitemShut
  {NoStop}%
\bibitem [{\citenamefont {Hassa{\"{i}}ne}\ and\ \citenamefont
  {Mart{\'{i}}nez}(2007)}]{Hassaine:2007py}%
  \BibitemOpen
  \bibfield  {author} {\bibinfo {author} {\bibfnamefont {M.}~\bibnamefont
  {Hassa{\"{i}}ne}}\ and\ \bibinfo {author} {\bibfnamefont {C.}~\bibnamefont
  {Mart{\'{i}}nez}},\ }\href {https://doi.org/10.1103/PhysRevD.75.027502}
  {\bibfield  {journal} {\bibinfo  {journal} {Phys. Rev. D}\ }\textbf {\bibinfo
  {volume} {75}},\ \bibinfo {pages} {027502} (\bibinfo {year} {2007})},\
  \Eprint {https://arxiv.org/abs/hep-th/0701058} {arXiv:hep-th/0701058}
  \BibitemShut {NoStop}%
\bibitem [{\citenamefont {C{\'{a}}rdenas}\ \emph {et~al.}(2014)\citenamefont
  {C{\'{a}}rdenas}, \citenamefont {Fuentealba},\ and\ \citenamefont
  {Mart\'\i{}nez}}]{Cardenas:2014kaa}%
  \BibitemOpen
  \bibfield  {author} {\bibinfo {author} {\bibfnamefont {M.}~\bibnamefont
  {C{\'{a}}rdenas}}, \bibinfo {author} {\bibfnamefont {O.}~\bibnamefont
  {Fuentealba}},\ and\ \bibinfo {author} {\bibfnamefont {C.}~\bibnamefont
  {Mart\'\i{}nez}},\ }\href {https://doi.org/10.1103/PhysRevD.90.124072}
  {\bibfield  {journal} {\bibinfo  {journal} {Phys. Rev. D}\ }\textbf {\bibinfo
  {volume} {90}},\ \bibinfo {pages} {124072} (\bibinfo {year} {2014})},\
  \Eprint {https://arxiv.org/abs/1408.1401} {arXiv:1408.1401 [hep-th]}
  \BibitemShut {NoStop}%
\bibitem [{\citenamefont {Ay{\'{o}}n-Beato}\ \emph {et~al.}(2006)\citenamefont
  {Ay{\'{o}}n-Beato}, \citenamefont {Mart{\'{i}}nez},\ and\ \citenamefont
  {Zanelli}}]{Ayon-Beato:2004nzi}%
  \BibitemOpen
  \bibfield  {author} {\bibinfo {author} {\bibfnamefont {E.}~\bibnamefont
  {Ay{\'{o}}n-Beato}}, \bibinfo {author} {\bibfnamefont {C.}~\bibnamefont
  {Mart{\'{i}}nez}},\ and\ \bibinfo {author} {\bibfnamefont {J.}~\bibnamefont
  {Zanelli}},\ }\href {https://doi.org/10.1007/s10714-005-0213-x} {\bibfield
  {journal} {\bibinfo  {journal} {Gen. Rel. Grav.}\ }\textbf {\bibinfo {volume}
  {38}},\ \bibinfo {pages} {145} (\bibinfo {year} {2006})},\ \Eprint
  {https://arxiv.org/abs/hep-th/0403228} {arXiv:hep-th/0403228} \BibitemShut
  {NoStop}%
\bibitem [{\citenamefont {Ay{\'{o}}n-Beato}\ \emph {et~al.}(2005)\citenamefont
  {Ay{\'{o}}n-Beato}, \citenamefont {Mart{\'{i}}nez}, \citenamefont
  {Troncoso},\ and\ \citenamefont {Zanelli}}]{Ayon-Beato:2005yoq}%
  \BibitemOpen
  \bibfield  {author} {\bibinfo {author} {\bibfnamefont {E.}~\bibnamefont
  {Ay{\'{o}}n-Beato}}, \bibinfo {author} {\bibfnamefont {C.}~\bibnamefont
  {Mart{\'{i}}nez}}, \bibinfo {author} {\bibfnamefont {R.}~\bibnamefont
  {Troncoso}},\ and\ \bibinfo {author} {\bibfnamefont {J.}~\bibnamefont
  {Zanelli}},\ }\href {https://doi.org/10.1103/PhysRevD.71.104037} {\bibfield
  {journal} {\bibinfo  {journal} {Phys. Rev. D}\ }\textbf {\bibinfo {volume}
  {71}},\ \bibinfo {pages} {104037} (\bibinfo {year} {2005})},\ \Eprint
  {https://arxiv.org/abs/hep-th/0505086} {arXiv:hep-th/0505086} \BibitemShut
  {NoStop}%
\bibitem [{\citenamefont {Ay\'on-Beato}\ \emph {et~al.}(2015)\citenamefont
  {Ay\'on-Beato}, \citenamefont {Hassa\"{\i}ne},\ and\ \citenamefont
  {Ju\'arez-Aubry}}]{Ayon-Beato:2015qfa}%
  \BibitemOpen
  \bibfield  {author} {\bibinfo {author} {\bibfnamefont {E.}~\bibnamefont
  {Ay\'on-Beato}}, \bibinfo {author} {\bibfnamefont {M.}~\bibnamefont
  {Hassa\"{\i}ne}},\ and\ \bibinfo {author} {\bibfnamefont {M.~M.}\
  \bibnamefont {Ju\'arez-Aubry}},\ }\href@noop {} {\bibinfo {title} {{Stealths
  on Anisotropic Holographic Backgrounds}}} (\bibinfo {year} {2015}),\ \Eprint
  {https://arxiv.org/abs/1506.03545} {arXiv:1506.03545 [gr-qc]} \BibitemShut
  {NoStop}%
\bibitem [{\citenamefont {Ay\'on-Beato}\ \emph {et~al.}(2018)\citenamefont
  {Ay\'on-Beato}, \citenamefont {Ram\'\i{}rez-Baca},\ and\ \citenamefont
  {Terrero-Escalante}}]{Ayon-Beato:2015mxf}%
  \BibitemOpen
  \bibfield  {author} {\bibinfo {author} {\bibfnamefont {E.}~\bibnamefont
  {Ay\'on-Beato}}, \bibinfo {author} {\bibfnamefont {P.~I.}\ \bibnamefont
  {Ram\'\i{}rez-Baca}},\ and\ \bibinfo {author} {\bibfnamefont {C.~A.}\
  \bibnamefont {Terrero-Escalante}},\ }\href
  {https://doi.org/10.1103/PhysRevD.97.043505} {\bibfield  {journal} {\bibinfo
  {journal} {Phys. Rev. D}\ }\textbf {\bibinfo {volume} {97}},\ \bibinfo
  {pages} {043505} (\bibinfo {year} {2018})},\ \Eprint
  {https://arxiv.org/abs/1512.09375} {arXiv:1512.09375 [gr-qc]} \BibitemShut
  {NoStop}%
\bibitem [{\citenamefont {Ay\'on-Beato}\ \emph {et~al.}(2024)\citenamefont
  {Ay\'on-Beato}, \citenamefont {Hassaine},\ and\ \citenamefont
  {S\'anchez}}]{Ayon-Beato:2024xgp}%
  \BibitemOpen
  \bibfield  {author} {\bibinfo {author} {\bibfnamefont {E.}~\bibnamefont
  {Ay\'on-Beato}}, \bibinfo {author} {\bibfnamefont {M.}~\bibnamefont
  {Hassaine}},\ and\ \bibinfo {author} {\bibfnamefont {P.~A.}\ \bibnamefont
  {S\'anchez}},\ }\href@noop {} {\bibinfo {title} {{Non-Noetherian Conformal
  Cheshire Effect}}} (\bibinfo {year} {2024}),\ \Eprint
  {https://arxiv.org/abs/2408.00086} {arXiv:2408.00086 [hep-th]} \BibitemShut
  {NoStop}%
\bibitem [{\citenamefont {Ay\'on-Beato}\ and\ \citenamefont
  {Hassa{\"{i}}ne}(2024)}]{Ayon-Beato:2023bzp}%
  \BibitemOpen
  \bibfield  {author} {\bibinfo {author} {\bibfnamefont {E.}~\bibnamefont
  {Ay\'on-Beato}}\ and\ \bibinfo {author} {\bibfnamefont {M.}~\bibnamefont
  {Hassa{\"{i}}ne}},\ }\href {https://doi.org/10.1016/j.aop.2023.169567}
  {\bibfield  {journal} {\bibinfo  {journal} {Annals Phys.}\ }\textbf {\bibinfo
  {volume} {460}},\ \bibinfo {pages} {169567} (\bibinfo {year} {2024})},\
  \Eprint {https://arxiv.org/abs/2305.09806} {arXiv:2305.09806 [hep-th]}
  \BibitemShut {NoStop}%
\bibitem [{\citenamefont {Fernandes}(2021)}]{Fernandes:2021dsb}%
  \BibitemOpen
  \bibfield  {author} {\bibinfo {author} {\bibfnamefont {P.~G.~S.}\
  \bibnamefont {Fernandes}},\ }\href
  {https://doi.org/10.1103/PhysRevD.103.104065} {\bibfield  {journal} {\bibinfo
   {journal} {Phys. Rev. D}\ }\textbf {\bibinfo {volume} {103}},\ \bibinfo
  {pages} {104065} (\bibinfo {year} {2021})},\ \Eprint
  {https://arxiv.org/abs/2105.04687} {arXiv:2105.04687 [gr-qc]} \BibitemShut
  {NoStop}%
\bibitem [{\citenamefont {Bocharova}\ \emph {et~al.}(1970)\citenamefont
  {Bocharova}, \citenamefont {Bronnikov},\ and\ \citenamefont
  {Melnikov}}]{Bocharova:1970skc}%
  \BibitemOpen
  \bibfield  {author} {\bibinfo {author} {\bibfnamefont {N.~M.}\ \bibnamefont
  {Bocharova}}, \bibinfo {author} {\bibfnamefont {K.~A.}\ \bibnamefont
  {Bronnikov}},\ and\ \bibinfo {author} {\bibfnamefont {V.~N.}\ \bibnamefont
  {Melnikov}},\ }\href {http://vmu.phys.msu.ru/file/1970/6/1970-6-11-706.pdf}
  {\bibfield  {journal} {\bibinfo  {journal} {Vestnik MGU Fiz. Astron. No. 6}\
  ,\ \bibinfo {pages} {706}} (\bibinfo {year} {1970})}\BibitemShut {NoStop}%
\bibitem [{\citenamefont {Bekenstein}(1974)}]{Bekenstein:1974sf}%
  \BibitemOpen
  \bibfield  {author} {\bibinfo {author} {\bibfnamefont {J.~D.}\ \bibnamefont
  {Bekenstein}},\ }\href {https://doi.org/10.1016/0003-4916(74)90124-9}
  {\bibfield  {journal} {\bibinfo  {journal} {Annals Phys.}\ }\textbf {\bibinfo
  {volume} {82}},\ \bibinfo {pages} {535} (\bibinfo {year} {1974})}\BibitemShut
  {NoStop}%
\bibitem [{\citenamefont {Sudarsky}\ and\ \citenamefont
  {Zannias}(1998)}]{Sudarsky:1997te}%
  \BibitemOpen
  \bibfield  {author} {\bibinfo {author} {\bibfnamefont {D.}~\bibnamefont
  {Sudarsky}}\ and\ \bibinfo {author} {\bibfnamefont {T.}~\bibnamefont
  {Zannias}},\ }\href {https://doi.org/10.1103/PhysRevD.58.087502} {\bibfield
  {journal} {\bibinfo  {journal} {Phys. Rev. D}\ }\textbf {\bibinfo {volume}
  {58}},\ \bibinfo {pages} {087502} (\bibinfo {year} {1998})},\ \Eprint
  {https://arxiv.org/abs/gr-qc/9712083} {arXiv:gr-qc/9712083} \BibitemShut
  {NoStop}%
\bibitem [{\citenamefont {Mart{\'{i}}nez}\ \emph {et~al.}(2003)\citenamefont
  {Mart{\'{i}}nez}, \citenamefont {Troncoso},\ and\ \citenamefont
  {Zanelli}}]{Martinez:2002ru}%
  \BibitemOpen
  \bibfield  {author} {\bibinfo {author} {\bibfnamefont {C.}~\bibnamefont
  {Mart{\'{i}}nez}}, \bibinfo {author} {\bibfnamefont {R.}~\bibnamefont
  {Troncoso}},\ and\ \bibinfo {author} {\bibfnamefont {J.}~\bibnamefont
  {Zanelli}},\ }\href {https://doi.org/10.1103/PhysRevD.67.024008} {\bibfield
  {journal} {\bibinfo  {journal} {Phys. Rev. D}\ }\textbf {\bibinfo {volume}
  {67}},\ \bibinfo {pages} {024008} (\bibinfo {year} {2003})},\ \Eprint
  {https://arxiv.org/abs/hep-th/0205319} {arXiv:hep-th/0205319} \BibitemShut
  {NoStop}%
\bibitem [{\citenamefont {Mart{\'{i}}nez}\ \emph {et~al.}(2006)\citenamefont
  {Mart{\'{i}}nez}, \citenamefont {Staforelli},\ and\ \citenamefont
  {Troncoso}}]{Martinez:2005di}%
  \BibitemOpen
  \bibfield  {author} {\bibinfo {author} {\bibfnamefont {C.}~\bibnamefont
  {Mart{\'{i}}nez}}, \bibinfo {author} {\bibfnamefont {J.~P.}\ \bibnamefont
  {Staforelli}},\ and\ \bibinfo {author} {\bibfnamefont {R.}~\bibnamefont
  {Troncoso}},\ }\href {https://doi.org/10.1103/PhysRevD.74.044028} {\bibfield
  {journal} {\bibinfo  {journal} {Phys. Rev. D}\ }\textbf {\bibinfo {volume}
  {74}},\ \bibinfo {pages} {044028} (\bibinfo {year} {2006})},\ \Eprint
  {https://arxiv.org/abs/hep-th/0512022} {arXiv:hep-th/0512022} \BibitemShut
  {NoStop}%
\bibitem [{\citenamefont {Anabal{\'{o}}n}\ and\ \citenamefont
  {Cisterna}(2012)}]{Anabalon:2012tu}%
  \BibitemOpen
  \bibfield  {author} {\bibinfo {author} {\bibfnamefont {A.}~\bibnamefont
  {Anabal{\'{o}}n}}\ and\ \bibinfo {author} {\bibfnamefont {A.}~\bibnamefont
  {Cisterna}},\ }\href {https://doi.org/10.1103/PhysRevD.85.084035} {\bibfield
  {journal} {\bibinfo  {journal} {Phys. Rev. D}\ }\textbf {\bibinfo {volume}
  {85}},\ \bibinfo {pages} {084035} (\bibinfo {year} {2012})},\ \Eprint
  {https://arxiv.org/abs/1201.2008} {arXiv:1201.2008 [hep-th]} \BibitemShut
  {NoStop}%
\bibitem [{\citenamefont {Bardoux}\ \emph {et~al.}(2014)\citenamefont
  {Bardoux}, \citenamefont {Caldarelli},\ and\ \citenamefont
  {Charmousis}}]{Bardoux:2013swa}%
  \BibitemOpen
  \bibfield  {author} {\bibinfo {author} {\bibfnamefont {Y.}~\bibnamefont
  {Bardoux}}, \bibinfo {author} {\bibfnamefont {M.~M.}\ \bibnamefont
  {Caldarelli}},\ and\ \bibinfo {author} {\bibfnamefont {C.}~\bibnamefont
  {Charmousis}},\ }\href {https://doi.org/10.1007/JHEP05(2014)039} {\bibfield
  {journal} {\bibinfo  {journal} {J. High Energy Phys.}\ }\textbf {\bibinfo
  {volume} {05}}\bibfield  {number} {\bibinfo  {number} { (2014)},\ \bibinfo
  {pages} {039}},\ }\Eprint {https://arxiv.org/abs/1311.1192} {arXiv:1311.1192
  [hep-th]} \BibitemShut {NoStop}%
\bibitem [{\citenamefont {Ay{\'{o}}n-Beato}\ \emph {et~al.}(2015)\citenamefont
  {Ay{\'{o}}n-Beato}, \citenamefont {Hassa{\"{i}}ne},\ and\ \citenamefont
  {M\'endez-Zavaleta}}]{Ayon-Beato:2015ada}%
  \BibitemOpen
  \bibfield  {author} {\bibinfo {author} {\bibfnamefont {E.}~\bibnamefont
  {Ay{\'{o}}n-Beato}}, \bibinfo {author} {\bibfnamefont {M.}~\bibnamefont
  {Hassa{\"{i}}ne}},\ and\ \bibinfo {author} {\bibfnamefont {J.~A.}\
  \bibnamefont {M\'endez-Zavaleta}},\ }\href
  {https://doi.org/10.1103/PhysRevD.92.024048} {\bibfield  {journal} {\bibinfo
  {journal} {Phys. Rev. D}\ }\textbf {\bibinfo {volume} {92}},\ \bibinfo
  {pages} {024048} (\bibinfo {year} {2015})},\ \Eprint
  {https://arxiv.org/abs/1506.02277} {arXiv:1506.02277 [hep-th]} \BibitemShut
  {NoStop}%
\bibitem [{\citenamefont {Barrientos}\ \emph
  {et~al.}(2022{\natexlab{a}})\citenamefont {Barrientos}, \citenamefont
  {Cisterna}, \citenamefont {Mora},\ and\ \citenamefont
  {Vigan\`o}}]{Barrientos:2022avi}%
  \BibitemOpen
  \bibfield  {author} {\bibinfo {author} {\bibfnamefont {J.}~\bibnamefont
  {Barrientos}}, \bibinfo {author} {\bibfnamefont {A.}~\bibnamefont
  {Cisterna}}, \bibinfo {author} {\bibfnamefont {N.}~\bibnamefont {Mora}},\
  and\ \bibinfo {author} {\bibfnamefont {A.}~\bibnamefont {Vigan\`o}},\ }\href
  {https://doi.org/10.1103/PhysRevD.106.024038} {\bibfield  {journal} {\bibinfo
   {journal} {Phys. Rev. D}\ }\textbf {\bibinfo {volume} {106}},\ \bibinfo
  {pages} {024038} (\bibinfo {year} {2022}{\natexlab{a}})},\ \Eprint
  {https://arxiv.org/abs/2202.06706} {arXiv:2202.06706 [hep-th]} \BibitemShut
  {NoStop}%
\bibitem [{\citenamefont {Bandos}\ \emph {et~al.}(2020)\citenamefont {Bandos},
  \citenamefont {Lechner}, \citenamefont {Sorokin},\ and\ \citenamefont
  {Townsend}}]{Bandos:2020jsw}%
  \BibitemOpen
  \bibfield  {author} {\bibinfo {author} {\bibfnamefont {I.}~\bibnamefont
  {Bandos}}, \bibinfo {author} {\bibfnamefont {K.}~\bibnamefont {Lechner}},
  \bibinfo {author} {\bibfnamefont {D.}~\bibnamefont {Sorokin}},\ and\ \bibinfo
  {author} {\bibfnamefont {P.~K.}\ \bibnamefont {Townsend}},\ }\href
  {https://doi.org/10.1103/PhysRevD.102.121703} {\bibfield  {journal} {\bibinfo
   {journal} {Phys. Rev. D}\ }\textbf {\bibinfo {volume} {102}},\ \bibinfo
  {pages} {121703} (\bibinfo {year} {2020})},\ \Eprint
  {https://arxiv.org/abs/2007.09092} {arXiv:2007.09092 [hep-th]} \BibitemShut
  {NoStop}%
\bibitem [{\citenamefont {Salazar}\ \emph {et~al.}(1987)\citenamefont
  {Salazar}, \citenamefont {Garc{\'{i}}a},\ and\ \citenamefont
  {Pleba{\'{n}}ski}}]{Salazar:1987ap}%
  \BibitemOpen
  \bibfield  {author} {\bibinfo {author} {\bibfnamefont {I.~H.}\ \bibnamefont
  {Salazar}}, \bibinfo {author} {\bibfnamefont {A.}~\bibnamefont
  {Garc{\'{i}}a}},\ and\ \bibinfo {author} {\bibfnamefont {J.}~\bibnamefont
  {Pleba{\'{n}}ski}},\ }\href {https://doi.org/10.1063/1.527430} {\bibfield
  {journal} {\bibinfo  {journal} {J. Math. Phys.}\ }\textbf {\bibinfo {volume}
  {28}},\ \bibinfo {pages} {2171} (\bibinfo {year} {1987})}\BibitemShut
  {NoStop}%
\bibitem [{\citenamefont {Pleba{\'{n}}ski}(1970)}]{Plebanski:1970}%
  \BibitemOpen
  \bibfield  {author} {\bibinfo {author} {\bibfnamefont {J.}~\bibnamefont
  {Pleba{\'{n}}ski}},\ }\href@noop {} {\emph {\bibinfo {title} {{Lectures on
  Non-Linear Electrodynamics}}}}\ (\bibinfo  {publisher} {Nordita},\ \bibinfo
  {address} {Copenhagen},\ \bibinfo {year} {1970})\BibitemShut {NoStop}%
\bibitem [{\citenamefont
  {Garc{\'{i}}a-D{\'{i}}az}(2021)}]{Garcia-Diaz:2021bao}%
  \BibitemOpen
  \bibfield  {author} {\bibinfo {author} {\bibfnamefont {A.~A.}\ \bibnamefont
  {Garc{\'{i}}a-D{\'{i}}az}},\ }\href@noop {} {\bibinfo {title} {{Stationary
  Rotating Black Hole Exact Solution within Einstein--Nonlinear
  Electrodynamics}}} (\bibinfo {year} {2021}),\ \Eprint
  {https://arxiv.org/abs/2112.06302} {arXiv:2112.06302 [gr-qc]} \BibitemShut
  {NoStop}%
\bibitem [{\citenamefont
  {Garc{\'{i}}a-D{\'{i}}az}(2022)}]{Garcia-Diaz:2022jpc}%
  \BibitemOpen
  \bibfield  {author} {\bibinfo {author} {\bibfnamefont {A.~A.}\ \bibnamefont
  {Garc{\'{i}}a-D{\'{i}}az}},\ }\href
  {https://doi.org/10.1016/j.aop.2022.168880} {\bibfield  {journal} {\bibinfo
  {journal} {Annals Phys.}\ }\textbf {\bibinfo {volume} {441}},\ \bibinfo
  {pages} {168880} (\bibinfo {year} {2022})},\ \Eprint
  {https://arxiv.org/abs/2201.10682} {arXiv:2201.10682 [gr-qc]} \BibitemShut
  {NoStop}%
\bibitem [{\citenamefont {Ay\'on-Beato}(2024)}]{Ayon-Beato:2022dwg}%
  \BibitemOpen
  \bibfield  {author} {\bibinfo {author} {\bibfnamefont {E.}~\bibnamefont
  {Ay\'on-Beato}},\ }\href {https://doi.org/10.1016/j.aop.2024.169771}
  {\bibfield  {journal} {\bibinfo  {journal} {Annals Phys.}\ }\textbf {\bibinfo
  {volume} {469}},\ \bibinfo {pages} {169771} (\bibinfo {year} {2024})},\
  \Eprint {https://arxiv.org/abs/2203.12809} {arXiv:2203.12809 [gr-qc]}
  \BibitemShut {NoStop}%
\bibitem [{\citenamefont {Kosyakov}(2020)}]{Kosyakov:2020wxv}%
  \BibitemOpen
  \bibfield  {author} {\bibinfo {author} {\bibfnamefont {B.~P.}\ \bibnamefont
  {Kosyakov}},\ }\href {https://doi.org/10.1016/j.physletb.2020.135840}
  {\bibfield  {journal} {\bibinfo  {journal} {Phys. Lett. B}\ }\textbf
  {\bibinfo {volume} {810}},\ \bibinfo {pages} {135840} (\bibinfo {year}
  {2020})},\ \Eprint {https://arxiv.org/abs/2007.13878} {arXiv:2007.13878
  [hep-th]} \BibitemShut {NoStop}%
\bibitem [{\citenamefont {Russo}\ and\ \citenamefont
  {Townsend}(2024)}]{Russo:2024llm}%
  \BibitemOpen
  \bibfield  {author} {\bibinfo {author} {\bibfnamefont {J.~G.}\ \bibnamefont
  {Russo}}\ and\ \bibinfo {author} {\bibfnamefont {P.~K.}\ \bibnamefont
  {Townsend}},\ }\href {https://doi.org/10.1103/PhysRevD.109.105023} {\bibfield
   {journal} {\bibinfo  {journal} {Phys. Rev. D}\ }\textbf {\bibinfo {volume}
  {109}},\ \bibinfo {pages} {105023} (\bibinfo {year} {2024})},\ \Eprint
  {https://arxiv.org/abs/2401.06707} {arXiv:2401.06707 [hep-th]} \BibitemShut
  {NoStop}%
\bibitem [{\citenamefont {Gibbons}\ and\ \citenamefont
  {Rasheed}(1995)}]{Gibbons:1995cv}%
  \BibitemOpen
  \bibfield  {author} {\bibinfo {author} {\bibfnamefont {G.~W.}\ \bibnamefont
  {Gibbons}}\ and\ \bibinfo {author} {\bibfnamefont {D.~A.}\ \bibnamefont
  {Rasheed}},\ }\href {https://doi.org/10.1016/0550-3213(95)00409-L} {\bibfield
   {journal} {\bibinfo  {journal} {Nucl. Phys. B}\ }\textbf {\bibinfo {volume}
  {454}},\ \bibinfo {pages} {185} (\bibinfo {year} {1995})},\ \Eprint
  {https://arxiv.org/abs/hep-th/9506035} {arXiv:hep-th/9506035} \BibitemShut
  {NoStop}%
\bibitem [{\citenamefont {Ballon~Bordo}\ \emph {et~al.}(2021)\citenamefont
  {Ballon~Bordo}, \citenamefont {Kubiz\v{n}\'ak},\ and\ \citenamefont
  {Perche}}]{BallonBordo:2020jtw}%
  \BibitemOpen
  \bibfield  {author} {\bibinfo {author} {\bibfnamefont {A.}~\bibnamefont
  {Ballon~Bordo}}, \bibinfo {author} {\bibfnamefont {D.}~\bibnamefont
  {Kubiz\v{n}\'ak}},\ and\ \bibinfo {author} {\bibfnamefont {T.~R.}\
  \bibnamefont {Perche}},\ }\href
  {https://doi.org/10.1016/j.physletb.2021.136312} {\bibfield  {journal}
  {\bibinfo  {journal} {Phys. Lett. B}\ }\textbf {\bibinfo {volume} {817}},\
  \bibinfo {pages} {136312} (\bibinfo {year} {2021})},\ \Eprint
  {https://arxiv.org/abs/2011.13398} {arXiv:2011.13398 [hep-th]} \BibitemShut
  {NoStop}%
\bibitem [{\citenamefont {Flores-Alfonso}\ \emph
  {et~al.}(2021{\natexlab{a}})\citenamefont {Flores-Alfonso}, \citenamefont
  {Linares},\ and\ \citenamefont {Maceda}}]{Flores-Alfonso:2020nnd}%
  \BibitemOpen
  \bibfield  {author} {\bibinfo {author} {\bibfnamefont {D.}~\bibnamefont
  {Flores-Alfonso}}, \bibinfo {author} {\bibfnamefont {R.}~\bibnamefont
  {Linares}},\ and\ \bibinfo {author} {\bibfnamefont {M.}~\bibnamefont
  {Maceda}},\ }\href {https://doi.org/10.1007/JHEP09(2021)104} {\bibfield
  {journal} {\bibinfo  {journal} {J. High Energy Phys.}\ }\textbf {\bibinfo
  {volume} {09}}\bibfield  {number} {\bibinfo  {number} { (2021)},\ \bibinfo
  {pages} {104}},\ }\Eprint {https://arxiv.org/abs/2012.03416}
  {arXiv:2012.03416 [gr-qc]} \BibitemShut {NoStop}%
\bibitem [{\citenamefont {Zhang}\ and\ \citenamefont
  {Jiang}(2021)}]{Zhang:2021qga}%
  \BibitemOpen
  \bibfield  {author} {\bibinfo {author} {\bibfnamefont {M.}~\bibnamefont
  {Zhang}}\ and\ \bibinfo {author} {\bibfnamefont {J.}~\bibnamefont {Jiang}},\
  }\href {https://doi.org/10.1103/PhysRevD.104.084094} {\bibfield  {journal}
  {\bibinfo  {journal} {Phys. Rev. D}\ }\textbf {\bibinfo {volume} {104}},\
  \bibinfo {pages} {084094} (\bibinfo {year} {2021})},\ \Eprint
  {https://arxiv.org/abs/2110.04757} {arXiv:2110.04757 [hep-th]} \BibitemShut
  {NoStop}%
\bibitem [{\citenamefont {Taub}(1951)}]{Taub:1950ez}%
  \BibitemOpen
  \bibfield  {author} {\bibinfo {author} {\bibfnamefont {A.~H.}\ \bibnamefont
  {Taub}},\ }\href {https://doi.org/10.2307/1969567} {\bibfield  {journal}
  {\bibinfo  {journal} {Annals Math.}\ }\textbf {\bibinfo {volume} {53}},\
  \bibinfo {pages} {472} (\bibinfo {year} {1951})}\BibitemShut {NoStop}%
\bibitem [{\citenamefont {Newman}\ \emph {et~al.}(1963)\citenamefont {Newman},
  \citenamefont {Tamburino},\ and\ \citenamefont {Unti}}]{Newman:1963yy}%
  \BibitemOpen
  \bibfield  {author} {\bibinfo {author} {\bibfnamefont {E.}~\bibnamefont
  {Newman}}, \bibinfo {author} {\bibfnamefont {L.}~\bibnamefont {Tamburino}},\
  and\ \bibinfo {author} {\bibfnamefont {T.}~\bibnamefont {Unti}},\ }\href
  {https://doi.org/10.1063/1.1704018} {\bibfield  {journal} {\bibinfo
  {journal} {J. Math. Phys.}\ }\textbf {\bibinfo {volume} {4}},\ \bibinfo
  {pages} {915} (\bibinfo {year} {1963})}\BibitemShut {NoStop}%
\bibitem [{\citenamefont {Misner}(1963)}]{Misner:1963fr}%
  \BibitemOpen
  \bibfield  {author} {\bibinfo {author} {\bibfnamefont {C.~W.}\ \bibnamefont
  {Misner}},\ }\href {https://doi.org/10.1063/1.1704019} {\bibfield  {journal}
  {\bibinfo  {journal} {J. Math. Phys.}\ }\textbf {\bibinfo {volume} {4}},\
  \bibinfo {pages} {924} (\bibinfo {year} {1963})}\BibitemShut {NoStop}%
\bibitem [{\citenamefont {Kim}\ \emph {et~al.}(2013)\citenamefont {Kim},
  \citenamefont {Kulkarni},\ and\ \citenamefont {Yi}}]{Kim:2013zha}%
  \BibitemOpen
  \bibfield  {author} {\bibinfo {author} {\bibfnamefont {W.}~\bibnamefont
  {Kim}}, \bibinfo {author} {\bibfnamefont {S.}~\bibnamefont {Kulkarni}},\ and\
  \bibinfo {author} {\bibfnamefont {S.-H.}\ \bibnamefont {Yi}},\ }\href
  {https://doi.org/10.1103/PhysRevLett.111.081101} {\bibfield  {journal}
  {\bibinfo  {journal} {Phys. Rev. Lett.}\ }\textbf {\bibinfo {volume} {111}},\
  \bibinfo {pages} {081101} (\bibinfo {year} {2013})},\ \bibinfo {note}
  {[Erratum: Phys.Rev.Lett. 112, 079902 (2014)]},\ \Eprint
  {https://arxiv.org/abs/1306.2138} {arXiv:1306.2138 [hep-th]} \BibitemShut
  {NoStop}%
\bibitem [{\citenamefont {Ay\'on-Beato}\ \emph {et~al.}(2015)\citenamefont
  {Ay\'on-Beato}, \citenamefont {Bravo-Gaete}, \citenamefont {Correa},
  \citenamefont {Hassa\"{\i}ne}, \citenamefont {Ju\'arez-Aubry},\ and\
  \citenamefont {Oliva}}]{Ayon-Beato:2015jga}%
  \BibitemOpen
  \bibfield  {author} {\bibinfo {author} {\bibfnamefont {E.}~\bibnamefont
  {Ay\'on-Beato}}, \bibinfo {author} {\bibfnamefont {M.}~\bibnamefont
  {Bravo-Gaete}}, \bibinfo {author} {\bibfnamefont {F.}~\bibnamefont {Correa}},
  \bibinfo {author} {\bibfnamefont {M.}~\bibnamefont {Hassa\"{\i}ne}}, \bibinfo
  {author} {\bibfnamefont {M.~M.}\ \bibnamefont {Ju\'arez-Aubry}},\ and\
  \bibinfo {author} {\bibfnamefont {J.}~\bibnamefont {Oliva}},\ }\href
  {https://doi.org/10.1103/PhysRevD.91.064006} {\bibfield  {journal} {\bibinfo
  {journal} {Phys. Rev. D}\ }\textbf {\bibinfo {volume} {91}},\ \bibinfo
  {pages} {064006} (\bibinfo {year} {2015})},\ \bibinfo {note} {[Addendum:
  Phys.Rev.D 96, 049903 (2017)]},\ \Eprint {https://arxiv.org/abs/1501.01244}
  {arXiv:1501.01244 [gr-qc]} \BibitemShut {NoStop}%
\bibitem [{\citenamefont {Ay\'on-Beato}\ \emph {et~al.}(2019)\citenamefont
  {Ay\'on-Beato}, \citenamefont {Bravo-Gaete}, \citenamefont {Correa},
  \citenamefont {Hassa{\"{i}}ne},\ and\ \citenamefont
  {Ju\'arez-Aubry}}]{Ayon-Beato:2019kmz}%
  \BibitemOpen
  \bibfield  {author} {\bibinfo {author} {\bibfnamefont {E.}~\bibnamefont
  {Ay\'on-Beato}}, \bibinfo {author} {\bibfnamefont {M.}~\bibnamefont
  {Bravo-Gaete}}, \bibinfo {author} {\bibfnamefont {F.}~\bibnamefont {Correa}},
  \bibinfo {author} {\bibfnamefont {M.}~\bibnamefont {Hassa{\"{i}}ne}},\ and\
  \bibinfo {author} {\bibfnamefont {M.~M.}\ \bibnamefont {Ju\'arez-Aubry}},\
  }\href {https://doi.org/10.1103/PhysRevD.100.044024} {\bibfield  {journal}
  {\bibinfo  {journal} {Phys. Rev. D}\ }\textbf {\bibinfo {volume} {100}},\
  \bibinfo {pages} {044024} (\bibinfo {year} {2019})},\ \Eprint
  {https://arxiv.org/abs/1904.09391} {arXiv:1904.09391 [hep-th]} \BibitemShut
  {NoStop}%
\bibitem [{\citenamefont {Barrientos}\ \emph
  {et~al.}(2022{\natexlab{b}})\citenamefont {Barrientos}, \citenamefont
  {Cisterna}, \citenamefont {Corral},\ and\ \citenamefont
  {Oyarzo}}]{Barrientos:2022yoz}%
  \BibitemOpen
  \bibfield  {author} {\bibinfo {author} {\bibfnamefont {J.}~\bibnamefont
  {Barrientos}}, \bibinfo {author} {\bibfnamefont {A.}~\bibnamefont
  {Cisterna}}, \bibinfo {author} {\bibfnamefont {C.}~\bibnamefont {Corral}},\
  and\ \bibinfo {author} {\bibfnamefont {M.}~\bibnamefont {Oyarzo}},\ }\href
  {https://doi.org/10.1007/JHEP05(2022)110} {\bibfield  {journal} {\bibinfo
  {journal} {J. High Energy Phys.}\ }\textbf {\bibinfo {volume} {05}}\bibfield
  {number} {\bibinfo  {number} { (2022)},\ \bibinfo {pages} {110}},\ }\Eprint
  {https://arxiv.org/abs/2202.13854} {arXiv:2202.13854 [hep-th]} \BibitemShut
  {NoStop}%
\bibitem [{\citenamefont {Colip\'\i{}-Marchant}\ \emph
  {et~al.}(2023)\citenamefont {Colip\'\i{}-Marchant}, \citenamefont {Corral},
  \citenamefont {Flores-Alfonso},\ and\ \citenamefont
  {Sanhueza}}]{Colipi-Marchant:2023awk}%
  \BibitemOpen
  \bibfield  {author} {\bibinfo {author} {\bibfnamefont {F.}~\bibnamefont
  {Colip\'\i{}-Marchant}}, \bibinfo {author} {\bibfnamefont {C.}~\bibnamefont
  {Corral}}, \bibinfo {author} {\bibfnamefont {D.}~\bibnamefont
  {Flores-Alfonso}},\ and\ \bibinfo {author} {\bibfnamefont {L.}~\bibnamefont
  {Sanhueza}},\ }\href {https://doi.org/10.1103/PhysRevD.107.104042} {\bibfield
   {journal} {\bibinfo  {journal} {Phys. Rev. D}\ }\textbf {\bibinfo {volume}
  {107}},\ \bibinfo {pages} {104042} (\bibinfo {year} {2023})},\ \Eprint
  {https://arxiv.org/abs/2302.09162} {arXiv:2302.09162 [hep-th]} \BibitemShut
  {NoStop}%
\bibitem [{\citenamefont {{Demia\'nski}}\ and\ \citenamefont
  {{Newman}}(1966)}]{Demianski:1966}%
  \BibitemOpen
  \bibfield  {author} {\bibinfo {author} {\bibfnamefont {M.}~\bibnamefont
  {{Demia\'nski}}}\ and\ \bibinfo {author} {\bibfnamefont {E.~T.}\ \bibnamefont
  {{Newman}}},\ }\href@noop {} {\bibfield  {journal} {\bibinfo  {journal}
  {Bull. Acad. Pol. Sci., Ser. Sci. Math. Astron. Phys.}\ }\textbf {\bibinfo
  {volume} {14}},\ \bibinfo {pages} {653} (\bibinfo {year} {1966})}\BibitemShut
  {NoStop}%
\bibitem [{\citenamefont {Flores-Alfonso}\ \emph
  {et~al.}(2021{\natexlab{b}})\citenamefont {Flores-Alfonso}, \citenamefont
  {Gonz\'alez-Morales}, \citenamefont {Linares},\ and\ \citenamefont
  {Maceda}}]{Flores-Alfonso:2020euz}%
  \BibitemOpen
  \bibfield  {author} {\bibinfo {author} {\bibfnamefont {D.}~\bibnamefont
  {Flores-Alfonso}}, \bibinfo {author} {\bibfnamefont {B.~A.}\ \bibnamefont
  {Gonz\'alez-Morales}}, \bibinfo {author} {\bibfnamefont {R.}~\bibnamefont
  {Linares}},\ and\ \bibinfo {author} {\bibfnamefont {M.}~\bibnamefont
  {Maceda}},\ }\href {https://doi.org/10.1016/j.physletb.2020.136011}
  {\bibfield  {journal} {\bibinfo  {journal} {Phys. Lett. B}\ }\textbf
  {\bibinfo {volume} {812}},\ \bibinfo {pages} {136011} (\bibinfo {year}
  {2021}{\natexlab{b}})},\ \Eprint {https://arxiv.org/abs/2011.10836}
  {arXiv:2011.10836 [gr-qc]} \BibitemShut {NoStop}%
\bibitem [{\citenamefont {Jackiw}(2006)}]{Jackiw:2005su}%
  \BibitemOpen
  \bibfield  {author} {\bibinfo {author} {\bibfnamefont {R.}~\bibnamefont
  {Jackiw}},\ }\href {https://doi.org/10.1007/s11232-006-0090-9} {\bibfield
  {journal} {\bibinfo  {journal} {Theor. Math. Phys.}\ }\textbf {\bibinfo
  {volume} {148}},\ \bibinfo {pages} {941} (\bibinfo {year} {2006})},\ \Eprint
  {https://arxiv.org/abs/hep-th/0511065} {arXiv:hep-th/0511065} \BibitemShut
  {NoStop}%
\bibitem [{\citenamefont {Ay\'on-Beato}\ and\ \citenamefont
  {Hassaine}(2023)}]{Ayon-Beato:2023lrn}%
  \BibitemOpen
  \bibfield  {author} {\bibinfo {author} {\bibfnamefont {E.}~\bibnamefont
  {Ay\'on-Beato}}\ and\ \bibinfo {author} {\bibfnamefont {M.}~\bibnamefont
  {Hassaine}},\ }\href {https://doi.org/10.1016/j.aop.2023.169446} {\bibfield
  {journal} {\bibinfo  {journal} {Annals Phys.}\ }\textbf {\bibinfo {volume}
  {458}},\ \bibinfo {pages} {169446} (\bibinfo {year} {2023})},\ \Eprint
  {https://arxiv.org/abs/2307.04048} {arXiv:2307.04048 [hep-th]} \BibitemShut
  {NoStop}%
\bibitem [{\citenamefont {Horndeski}(1974)}]{Horndeski:1974wa}%
  \BibitemOpen
  \bibfield  {author} {\bibinfo {author} {\bibfnamefont {G.~W.}\ \bibnamefont
  {Horndeski}},\ }\href {https://doi.org/10.1007/BF01807638} {\bibfield
  {journal} {\bibinfo  {journal} {Int. J. Theor. Phys.}\ }\textbf {\bibinfo
  {volume} {10}},\ \bibinfo {pages} {363} (\bibinfo {year} {1974})}\BibitemShut
  {NoStop}%
\bibitem [{\citenamefont {Babichev}\ \emph {et~al.}(2023)\citenamefont
  {Babichev}, \citenamefont {Charmousis}, \citenamefont {Hassa{\"{i}}ne},\ and\
  \citenamefont {Lecoeur}}]{Babichev:2023rhn}%
  \BibitemOpen
  \bibfield  {author} {\bibinfo {author} {\bibfnamefont {E.}~\bibnamefont
  {Babichev}}, \bibinfo {author} {\bibfnamefont {C.}~\bibnamefont
  {Charmousis}}, \bibinfo {author} {\bibfnamefont {M.}~\bibnamefont
  {Hassa{\"{i}}ne}},\ and\ \bibinfo {author} {\bibfnamefont {N.}~\bibnamefont
  {Lecoeur}},\ }\href {https://doi.org/10.1103/PhysRevD.107.084050} {\bibfield
  {journal} {\bibinfo  {journal} {Phys. Rev. D}\ }\textbf {\bibinfo {volume}
  {107}},\ \bibinfo {pages} {084050} (\bibinfo {year} {2023})},\ \Eprint
  {https://arxiv.org/abs/2302.02920} {arXiv:2302.02920 [gr-qc]} \BibitemShut
  {NoStop}%
\bibitem [{\citenamefont {Hassa{\"{i}}ne}\ \emph {et~al.}(2023)\citenamefont
  {Hassa{\"{i}}ne}, \citenamefont {Hernandez-Vera},\ and\ \citenamefont
  {Lara-Munoz}}]{Hassaine:2023paj}%
  \BibitemOpen
  \bibfield  {author} {\bibinfo {author} {\bibfnamefont {M.}~\bibnamefont
  {Hassa{\"{i}}ne}}, \bibinfo {author} {\bibfnamefont {U.}~\bibnamefont
  {Hernandez-Vera}},\ and\ \bibinfo {author} {\bibfnamefont {F.}~\bibnamefont
  {Lara-Munoz}},\ }\href {https://doi.org/10.1103/PhysRevD.108.104067}
  {\bibfield  {journal} {\bibinfo  {journal} {Phys. Rev. D}\ }\textbf {\bibinfo
  {volume} {108}},\ \bibinfo {pages} {104067} (\bibinfo {year} {2023})},\
  \Eprint {https://arxiv.org/abs/2309.03024} {arXiv:2309.03024 [hep-th]}
  \BibitemShut {NoStop}%
\bibitem [{\citenamefont
  {Bia{\l}ynicki-Birula}(1983)}]{Bialynicki-Birula:1984daz}%
  \BibitemOpen
  \bibfield  {author} {\bibinfo {author} {\bibfnamefont {I.}~\bibnamefont
  {Bia{\l}ynicki-Birula}},\ }\bibinfo {title} {{Nonlinear electrodynamics:
  variations on a theme by Born and Infeld}},\ in\ \href@noop {} {\emph
  {\bibinfo {booktitle} {Quantum Theory of Particles and Fields}}}\ (\bibinfo
  {publisher} {World Scientific},\ \bibinfo {address} {Singapore},\ \bibinfo
  {year} {1983})\ pp.\ \bibinfo {pages} {31--48}\BibitemShut {NoStop}%
\bibitem [{\citenamefont
  {Bia{\l}ynicki-Birula}(1992)}]{Bialynicki-Birula:1992rcm}%
  \BibitemOpen
  \bibfield  {author} {\bibinfo {author} {\bibfnamefont {I.}~\bibnamefont
  {Bia{\l}ynicki-Birula}},\ }\href
  {https://inspirehep.net/files/c22ef1b2bf81b30d8641572f4451a357} {\bibfield
  {journal} {\bibinfo  {journal} {Acta Phys. Polon. B}\ }\textbf {\bibinfo
  {volume} {23}},\ \bibinfo {pages} {553} (\bibinfo {year} {1992})}\BibitemShut
  {NoStop}%
\bibitem [{\citenamefont {Bravo-Gaete}\ \emph {et~al.}(2022)\citenamefont
  {Bravo-Gaete}, \citenamefont {Guajardo},\ and\ \citenamefont
  {Oliva}}]{Bravo-Gaete:2022mnr}%
  \BibitemOpen
  \bibfield  {author} {\bibinfo {author} {\bibfnamefont {M.}~\bibnamefont
  {Bravo-Gaete}}, \bibinfo {author} {\bibfnamefont {L.}~\bibnamefont
  {Guajardo}},\ and\ \bibinfo {author} {\bibfnamefont {J.}~\bibnamefont
  {Oliva}},\ }\href {https://doi.org/10.1103/PhysRevD.106.024017} {\bibfield
  {journal} {\bibinfo  {journal} {Phys. Rev. D}\ }\textbf {\bibinfo {volume}
  {106}},\ \bibinfo {pages} {024017} (\bibinfo {year} {2022})},\ \Eprint
  {https://arxiv.org/abs/2205.09282} {arXiv:2205.09282 [hep-th]} \BibitemShut
  {NoStop}%
\bibitem [{\citenamefont {Anastasiou}\ \emph {et~al.}(2023)\citenamefont
  {Anastasiou}, \citenamefont {Araya}, \citenamefont {Busnego-Barrientos},
  \citenamefont {Corral},\ and\ \citenamefont {Merino}}]{Anastasiou:2022wjq}%
  \BibitemOpen
  \bibfield  {author} {\bibinfo {author} {\bibfnamefont {G.}~\bibnamefont
  {Anastasiou}}, \bibinfo {author} {\bibfnamefont {I.~J.}\ \bibnamefont
  {Araya}}, \bibinfo {author} {\bibfnamefont {M.}~\bibnamefont
  {Busnego-Barrientos}}, \bibinfo {author} {\bibfnamefont {C.}~\bibnamefont
  {Corral}},\ and\ \bibinfo {author} {\bibfnamefont {N.}~\bibnamefont
  {Merino}},\ }\href {https://doi.org/10.1103/PhysRevD.107.104049} {\bibfield
  {journal} {\bibinfo  {journal} {Phys. Rev. D}\ }\textbf {\bibinfo {volume}
  {107}},\ \bibinfo {pages} {104049} (\bibinfo {year} {2023})},\ \Eprint
  {https://arxiv.org/abs/2212.04364} {arXiv:2212.04364 [hep-th]} \BibitemShut
  {NoStop}%
\bibitem [{\citenamefont {Podolsk{\'{y}}}\ and\ \citenamefont
  {Vr{\'{a}}tn{\'{y}}}(2021)}]{Podolsky:2021zwr}%
  \BibitemOpen
  \bibfield  {author} {\bibinfo {author} {\bibfnamefont {J.}~\bibnamefont
  {Podolsk{\'{y}}}}\ and\ \bibinfo {author} {\bibfnamefont {A.}~\bibnamefont
  {Vr{\'{a}}tn{\'{y}}}},\ }\href {https://doi.org/10.1103/PhysRevD.104.084078}
  {\bibfield  {journal} {\bibinfo  {journal} {Phys. Rev. D}\ }\textbf {\bibinfo
  {volume} {104}},\ \bibinfo {pages} {084078} (\bibinfo {year} {2021})},\
  \Eprint {https://arxiv.org/abs/2108.02239} {arXiv:2108.02239 [gr-qc]}
  \BibitemShut {NoStop}%
\bibitem [{\citenamefont {Podolsk{\'{y}}}\ and\ \citenamefont
  {Vr{\'{a}}tn{\'{y}}}(2023)}]{Podolsky:2022xxd}%
  \BibitemOpen
  \bibfield  {author} {\bibinfo {author} {\bibfnamefont {J.}~\bibnamefont
  {Podolsk{\'{y}}}}\ and\ \bibinfo {author} {\bibfnamefont {A.}~\bibnamefont
  {Vr{\'{a}}tn{\'{y}}}},\ }\href {https://doi.org/10.1103/PhysRevD.107.084034}
  {\bibfield  {journal} {\bibinfo  {journal} {Phys. Rev. D}\ }\textbf {\bibinfo
  {volume} {107}},\ \bibinfo {pages} {084034} (\bibinfo {year} {2023})},\
  \bibinfo {note} {[Erratum: Phys.Rev.D 108, 129902 (2023)]},\ \Eprint
  {https://arxiv.org/abs/2212.08865} {arXiv:2212.08865 [gr-qc]} \BibitemShut
  {NoStop}%
\bibitem [{\citenamefont {Astorino}\ and\ \citenamefont
  {Boldi}(2023)}]{Astorino:2023elf}%
  \BibitemOpen
  \bibfield  {author} {\bibinfo {author} {\bibfnamefont {M.}~\bibnamefont
  {Astorino}}\ and\ \bibinfo {author} {\bibfnamefont {G.}~\bibnamefont
  {Boldi}},\ }\href {https://doi.org/10.1007/JHEP08(2023)085} {\bibfield
  {journal} {\bibinfo  {journal} {J. High Energy Phys.}\ }\textbf {\bibinfo
  {volume} {08}}\bibfield  {number} {\bibinfo  {number} { (2023)},\ \bibinfo
  {pages} {085}},\ }\Eprint {https://arxiv.org/abs/2305.03744}
  {arXiv:2305.03744 [gr-qc]} \BibitemShut {NoStop}%
\bibitem [{\citenamefont {Astorino}(2024{\natexlab{a}})}]{Astorino:2023uim}%
  \BibitemOpen
  \bibfield  {author} {\bibinfo {author} {\bibfnamefont {M.}~\bibnamefont
  {Astorino}},\ }\href {https://doi.org/10.1103/PhysRevD.109.084038} {\bibfield
   {journal} {\bibinfo  {journal} {Phys. Rev. D}\ }\textbf {\bibinfo {volume}
  {109}},\ \bibinfo {pages} {084038} (\bibinfo {year} {2024}{\natexlab{a}})},\
  \Eprint {https://arxiv.org/abs/2312.00865} {arXiv:2312.00865 [gr-qc]}
  \BibitemShut {NoStop}%
\bibitem [{\citenamefont {Astorino}(2024{\natexlab{b}})}]{Astorino:2024bfl}%
  \BibitemOpen
  \bibfield  {author} {\bibinfo {author} {\bibfnamefont {M.}~\bibnamefont
  {Astorino}},\ }\href@noop {} {\bibinfo {title} {{Most general Type-D Black
  Hole and Accelerating Reissner-Nordstrom-NUT-(A)dS}}} (\bibinfo {year}
  {2024}{\natexlab{b}}),\ \Eprint {https://arxiv.org/abs/2404.06551}
  {arXiv:2404.06551 [gr-qc]} \BibitemShut {NoStop}%
\bibitem [{\citenamefont {Pleba{\'{n}}ski}\ and\ \citenamefont
  {Demia{\'{n}}ski}(1976)}]{Plebanski:1976gy}%
  \BibitemOpen
  \bibfield  {author} {\bibinfo {author} {\bibfnamefont {J.~F.}\ \bibnamefont
  {Pleba{\'{n}}ski}}\ and\ \bibinfo {author} {\bibfnamefont {M.}~\bibnamefont
  {Demia{\'{n}}ski}},\ }\href {https://doi.org/10.1016/0003-4916(76)90240-2}
  {\bibfield  {journal} {\bibinfo  {journal} {Annals Phys.}\ }\textbf {\bibinfo
  {volume} {98}},\ \bibinfo {pages} {98} (\bibinfo {year} {1976})}\BibitemShut
  {NoStop}%
\bibitem [{\citenamefont {Chng}\ \emph {et~al.}(2006)\citenamefont {Chng},
  \citenamefont {Mann},\ and\ \citenamefont {Stelea}}]{Chng:2006gh}%
  \BibitemOpen
  \bibfield  {author} {\bibinfo {author} {\bibfnamefont {B.}~\bibnamefont
  {Chng}}, \bibinfo {author} {\bibfnamefont {R.~B.}\ \bibnamefont {Mann}},\
  and\ \bibinfo {author} {\bibfnamefont {C.}~\bibnamefont {Stelea}},\ }\href
  {https://doi.org/10.1103/PhysRevD.74.084031} {\bibfield  {journal} {\bibinfo
  {journal} {Phys. Rev. D}\ }\textbf {\bibinfo {volume} {74}},\ \bibinfo
  {pages} {084031} (\bibinfo {year} {2006})},\ \Eprint
  {https://arxiv.org/abs/gr-qc/0608092} {arXiv:gr-qc/0608092} \BibitemShut
  {NoStop}%
\bibitem [{\citenamefont {Podolsk{\'{y}}}\ and\ \citenamefont
  {Vr{\'{a}}tn{\'{y}}}(2020)}]{Podolsky:2020xkf}%
  \BibitemOpen
  \bibfield  {author} {\bibinfo {author} {\bibfnamefont {J.}~\bibnamefont
  {Podolsk{\'{y}}}}\ and\ \bibinfo {author} {\bibfnamefont {A.}~\bibnamefont
  {Vr{\'{a}}tn{\'{y}}}},\ }\href {https://doi.org/10.1103/PhysRevD.102.084024}
  {\bibfield  {journal} {\bibinfo  {journal} {Phys. Rev. D}\ }\textbf {\bibinfo
  {volume} {102}},\ \bibinfo {pages} {084024} (\bibinfo {year} {2020})},\
  \Eprint {https://arxiv.org/abs/2007.09169} {arXiv:2007.09169 [gr-qc]}
  \BibitemShut {NoStop}%
\bibitem [{\citenamefont {Barrientos}\ and\ \citenamefont
  {Cisterna}(2023)}]{Barrientos:2023tqb}%
  \BibitemOpen
  \bibfield  {author} {\bibinfo {author} {\bibfnamefont {J.}~\bibnamefont
  {Barrientos}}\ and\ \bibinfo {author} {\bibfnamefont {A.}~\bibnamefont
  {Cisterna}},\ }\href {https://doi.org/10.1103/PhysRevD.108.024059} {\bibfield
   {journal} {\bibinfo  {journal} {Phys. Rev. D}\ }\textbf {\bibinfo {volume}
  {108}},\ \bibinfo {pages} {024059} (\bibinfo {year} {2023})},\ \Eprint
  {https://arxiv.org/abs/2305.03765} {arXiv:2305.03765 [gr-qc]} \BibitemShut
  {NoStop}%
\bibitem [{\citenamefont {Barrientos}\ \emph
  {et~al.}(2022{\natexlab{c}})\citenamefont {Barrientos}, \citenamefont
  {Cisterna}, \citenamefont {Kubiz{\v{n}}{\'a}k},\ and\ \citenamefont
  {Oliva}}]{Barrientos:2022bzm}%
  \BibitemOpen
  \bibfield  {author} {\bibinfo {author} {\bibfnamefont {J.}~\bibnamefont
  {Barrientos}}, \bibinfo {author} {\bibfnamefont {A.}~\bibnamefont
  {Cisterna}}, \bibinfo {author} {\bibfnamefont {D.}~\bibnamefont
  {Kubiz{\v{n}}{\'a}k}},\ and\ \bibinfo {author} {\bibfnamefont
  {J.}~\bibnamefont {Oliva}},\ }\href
  {https://doi.org/10.1016/j.physletb.2022.137447} {\bibfield  {journal}
  {\bibinfo  {journal} {Phys. Lett. B}\ }\textbf {\bibinfo {volume} {834}},\
  \bibinfo {pages} {137447} (\bibinfo {year} {2022}{\natexlab{c}})},\ \Eprint
  {https://arxiv.org/abs/2205.15777} {arXiv:2205.15777 [gr-qc]} \BibitemShut
  {NoStop}%
\end{thebibliography}
\end{document}